\documentclass[5p,times,nopreprintline]{elsarticle}

\usepackage{amsmath, amsthm, amssymb, amsfonts, amsbsy, mathrsfs}

\usepackage{xcolor}
\usepackage{calligra,bm}
\usepackage{stmaryrd}
\usepackage{yfonts}
\usepackage{accents}
\usepackage{multirow}
\usepackage{graphicx}
\usepackage{etoolbox}

\newcommand{\ubar}[1]{\underaccent{\bar}{#1}}

\usepackage[hidelinks,bookmarks=true]{hyperref}
\hypersetup{pdfstartview=FitH,pdfhighlight=/O,colorlinks=false}
\usepackage{orcidlink}

\begin{document}

\begin{frontmatter}

\title{\raggedright Redshifted civilizations, galactic empires, and the
                    Fermi paradox}

\author[inst1]{Chris Reiss}
\ead{christopher.j.reiss@gmail.com}
\address[inst1]{Independent Researcher}

\author[inst2]{Justin C. Feng\orcidlink{0000-0003-2441-5801}}
\ead{feng@fzu.cz}
\address[inst2]{Central European Institute for Cosmology and
    Fundamental Physics, Institute of Physics of the Czech Academy of
    Sciences, Na Slovance 1999/2, 182 21 Prague 8, Czech Republic}

%=======================================================================
%-----------------------------------------------------------------------
%
%		ABSTRACT
%
%-----------------------------------------------------------------------
%=======================================================================

\begin{abstract}
    Given the vast distances between stars in the Milky Way and the long
    timescales required for interstellar travel, we consider how a
    civilization might overcome the constraints arising from finite
    lifespans and the speed of light without invoking exotic or novel
    physics. We consider several scenarios in which a civilization can
    migrate to a time-dilated frame within the scope of classical
    general relativity and without incurring a biologically intolerable
    level of acceleration. Remarkably, the power requirements are lower
    than one might expect; biologically tolerable orbits near the photon
    radius of Sgr A* can be maintained by a civilization well below the
    Type II threshold, and a single Type II civilization can establish a
    galaxy-spanning civilization with a time dilation factor of $10^4$,
    enabling trips spanning the diameter of the Milky Way within a human
    lifetime in the civilizational reference frame. We also find that
    isotropic, monochromatic signals from orbits near the photon radius
    of a black hole exhibit a downward frequency drift. The
    vulnerability of ultrarelativistic vessels to destruction, combined
    with the relatively short timescales on which adversarial
    civilizations can arise, provides a strong motivating element for
    the ``dark forest'' hypothesis.
\end{abstract}

\end{frontmatter}

%-----------------------------------------------------------------------
%-----------------------------------
%-----------------
%--------
%---
%-
%
%
%-
%---
%--------
%-----------------
%-----------------------------------
%-----------------------------------------------------------------------

%=======================================================================
%-----------------------------------------------------------------------
%
%		INTRODUCTION
%
%-----------------------------------------------------------------------
%=======================================================================

\section{Introduction}

Given the human drive for exploration, how might a future human
civilization cope with the constraints arising from finite individual
lifespans and the speed of light? This question was explored by Robert
Goddard and Konstantin Tsiolkovsky \cite{Tsiolkovsky:1928} roughly a
century ago in a couple of essays \cite{Goddard:1918,Tsiolkovsky:1928}.
Goddard described passengers in suspended animation during interstellar
travel, and Tsiolkovsky imagined voyages comprised of multiple
generations of passengers. Both of these proposals present the same
severe drawback: the home civilization would be separated from the
voyagers for an extended period of time. Individuals on these voyages
might be separated from friends and family for decades, centuries, or
millennia, and even if human lifespans can be extended indefinitely,
such a long separation may incur a significant amount of emotional
stress (this is studied in the framework of attachment theory, see
\cite{Cassidy:2016} for an overview). 

These and related considerations perhaps motivate the significant body
of research into the possibility of faster-than-light travel. However,
faster-than-light travel requires either an enormous amount of exotic
matter or modifications to gravity
\cite{Alcubierre:1994tu,Visser1995,VanDenBroeck:1999sn,Lobo:2017cay}.
Regarding warp drives, it was pointed out in \cite{Coule:1998} that the
exotic matter in a warp drive must also be tachyonic, so that you need a
current of exotic matter to move faster than light in order to make a
warp drive. In both cases, new physics is required. It is perhaps
prudent to avoid speculating on new physics, and to explore
possibilities within the scope of classical general relativity (and the
known properties of matter), as it is a well-established theory on the
astrophysical scales of interest, from solar system tests
\cite{Will:2014kxa} to gravitational wave observations from the
collisions of compact objects (such as black holes)
\cite{Yunes:2013dva}.

In this article, we show that both special and general relativity offer
some possibilities for travelers (which we assume to have finite
lifespans) in a future human civilization to explore a significant
portion of the galaxy without invoking new or exotic physics and without
sacrificing for the possibility of reunion with individuals that remain
at home. Earlier proposals have exploited the relativistic phenomenon of
time dilation to enable travel over great distances within the lifespan
of the (human) traveler. For instance, Carl Sagan pointed out that at a
constant acceleration of $1~\mathrm{g}\approx 10~\mathrm{m/s}^2$, a
traveler can reach the galactic center (Sgr A) in $21~\mathrm{yr}$ in
ship time (the proper elapsed time experienced by the travelers), and to
Andromeda (M31) in $28~\mathrm{yr}$ in ship time \cite{Sagan:1963}.

Of course, an attempt to exploit time dilation in the manner described
by Sagan \cite{Sagan:1963} does not by itself solve the problem
described earlier, namely that of extreme separation times for long
distance voyages. The resolution of the so-called twin paradox in
special relativity indicates that extreme disparities arise in the
elapsed time experienced by travelers on round-trip voyages and the home
civilization \cite{TaylorWheeler:1992}. For instance, a terrestrial
traveler may travel to the galactic center and return to Earth,
experiencing only a few decades of aging during the voyage. But upon
returning, the traveler discovers that tens of \textit{millennia} have
passed on the Earth. One might imagine that the extreme temporal
disparities between traveler and the home civilization may become
unbearable.

We propose a rather simple way to avoid such temporal
disparities: simply put, we move the entire civilization to a highly
time-dilated frame. In particular, one might imagine scenarios in which
the civilization resides on vessels accelerated to ultrarelativistic
velocities. In this manner, travelers can chart courses that minimize
temporal disparities.  

Apart from a very preliminary exploration by one of the present authors
\cite{Reiss:2020} (which this article expands on greatly), it is rather
surprising that a detailed consideration of such scenarios does not yet
appear in the academic and scientific literature. It has, for instance,
been more than a decade since Ashworth's postulate (in a mostly
nonrelativistic context) that the dominant mode of human civilization
may shift from planetary to space based life
\cite{Ashworth:2012emergence}. We do note that others have considered
the effects of time dilation on trade, governance, and social dynamics
\cite{Krugman:2010theory,Matloff:2019motivation,Hang:2022social,Hang:2023interstellar,Besteiro:2019implications}.
Another idea considered in the literature is the possibility that
civilizations may exploit time dilation near black hole horizons to
``hibernate'' \cite{Dyson:1979,Vidal:2014,Vidal:2011} (see also
\cite{Li:2022}), but as we will discuss in some detail, a textbook
analysis of geodesics in the Schwarzschild geometry reveals that the
acceleration required to maintain a position just outside a black hole
is far beyond what biological organisms can tolerate.

One can instead consider orbits near the innermost stable circular orbit
of a supermassive black hole, in which observers can experience a
significant time dilation relative to those far away from the black
hole. This scenario was considered by Kip Thorne for Miller's planet in
the movie {\it Interstellar} \cite{Thorne:2014}; the time dilation
(Lorentz) factor of $\sim 6\times10^4$ considered requires a
supermassive black hole with a mass of $10^8~M_\odot$ with a spin that
differs from the extremal limit by one part in $100$ trillion
\cite{Thorne:2014}. While the mass of such a black hole falls within the
range of known supermassive black holes, such a finely-tuned spin
parameter is highly unrealistic, far from the theoretical limit
achievable by known astrophysical processes
\cite{Thorne:1974ve,Kesden:2009ds}. For comparison, the black hole at
the center of the Milky Way, Sgr A*, has a mass of
$4.3\times10^6~M_{\odot}$ \cite{GRAVITY:2023avo}, and a spin parameter
of $a=0.90 \pm 0.06$ \cite{Daly:2023axh}, and the recently imaged M87*
has a mass $6.5\times10^9~M_{\odot}$
\cite{EventHorizonTelescope:2019ggy} and spin parameter $a=0.90\pm0.05$
\cite{Daly:2023axh}. It should be mentioned that while it is rather
obvious that a civilization that migrates to Miller's planet can exploit
time dilation, no mention of this possibility is made in the movie or in
\cite{Thorne:2014}.

As indicated earlier, we consider how a civilization might migrate to a
time-dilated, or ``redshifted'' frame. As one might imagine, such a
migration would be an undertaking of enormous technical scope, requiring
the development of technologies that are perhaps currently beyond our
present ability to imagine, even if we limit our analysis to
well-established physics (cf. the remarks in Ch. 7 of
\cite{Davies:2010book}). For this reason, we refrain from making
detailed assumptions about the technological capabilities of such a
civilization, focusing on constraints arising from fundamental physics
and (human) biology. We take the perspective that the overall capability
of a civilization is mainly determined by raw power capacity rather than
specific technologies. In short, our analysis applies three fundamental
constraints:
\begin{itemize}
    \item[1.] Biological matter cannot tolerate high acceleration,
    \item[2.] General relativity is valid on scales of interest,
    \item[3.] Energy is nonnegative and conserved locally.
\end{itemize}
The reader may already be familiar with constraint 1, which may perhaps
be apparent from the significant number of annual fatalities from
automobile accidents \cite{Spencer:2021motor}; these provide a constant 
reminder of the intolerance of human biology to the mechanical stress 
from high acceleration. 
We have already discussed constraint 2, but 
since we discuss physics on galactic scales, the reader might be 
concerned about the possibility of modifications to gravitational 
dynamics at low accelerations 
\cite{Milgrom:1983zz,Milgrom:1983ca,Milgrom:1983pn,Bekenstein:1984tv} 
(on the order of $10^{-10}~\mathrm{m/s}^2$ \cite{Begeman:1991iy}) as an
alternative to dark matter, although the present evidence largely seems
to favor the latter as an explanation for galactic rotation curves
\cite{Balazs:2024uyj}. However, such modifications are largely irrelevant
for our analysis, which focuses on acceleration scales on the order of
$\sim 10~\mathrm{m/s}^2$ or greater. Finally, constraint 3 is reasonable
for the situations we consider; energy conservation holds exactly in
special relativity and also in the general relativistic spacetime
geometries describing isolated black holes (cf. Sec. 25.2--25.3 of
\cite{MTW}). This last constraint is useful for establishing limits for
the capabilities of a civilization based on its power capacity, that is,
its classification on the Kardashev scale \cite{Kardashev:1964}.

A few remarks on terminology and notation are in order. The ``galactic
frame'' will refer to the collection of reference frames associated with
the average velocity of stars in the galactic disk, which is less than a
thousandth of the speed of light---for our purposes, this velocity is
negligible, so we do not consider in detail the motion of stars in the
galactic disk. The ``civilizational frame'' refers to the collection of
reference frames associated with the vessels comprising the
civilizations we consider in this article. In this article, we use the
terms ``time-dilated'' and ``redshifted'' interchangeably. We treat the
term ``time dilation factor'' synonymously with ``Lorentz factor'',
except in curved spacetime. Greek letter superscripts and subscripts
denote spacetime tensor indices---the tensor formalism is used sparingly
in the main text, and we leave detailed tensor calculations in special
and general relativity in the appendix.

This paper is organized as follows. We begin with a discussion of the
phenomenon of time dilation in section \ref{sec:TimeDilation}. In
sections \ref{sec:CivSMBH}, \ref{Sec:LinearMotion}, and
\ref{Sec:RingBH}, we describe redshifted civilizations, and establish
constraints for these civilizations. In section \ref{sec:Fermi}, we
discuss implications for the Fermi paradox, and conclude with a summary
and some discussion of our results and their implications in section
\ref{sec:conc}. As mentioned in the preceding paragraph, detailed
technical calculations in special and general relativity are presented
in \ref{sec:AppendixGRSR}.

%=======================================================================
%-----------------------------------------------------------------------
%
%    Time dilation in special and general relativity
%
%-----------------------------------------------------------------------
%=======================================================================

\section{Time dilation in special and general
relativity}\label{sec:TimeDilation}

In this section, we review the phenomenon of time dilation in special
and general relativity. The phenomenon of time dilation has been
well-established in special and general relativity; the Hafele--Keating
experiment \cite{Hafele:1972aro} is a particularly well-known example,
but the phenomena has also been confirmed in particle physics
experiments
\cite{Bailey:1977mea,Bailey:1979fin,Roos:1980sig,Coan:2006com}.
Remarkably, gravitational time dilation has been measured in the
laboratory by independent teams over distances of a few millimeters
\cite{Bothwell:2022res,Zheng:2022dif}. Moreover, both special
relativistic and gravitational time dilation must be properly accounted
for in satellite navigation systems (the Global Positioning System and
the Galileo GNSS for instance) in order to avoid accumulated errors on
the order of $11~\mathrm{km}$ per day
\cite{Ashby:2002gps,Ashby:2003vja}.

\begin{figure}[htp]
    \centering
    \includegraphics[width=0.20\textwidth]{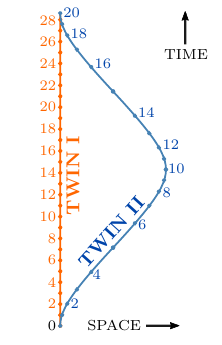}
    \caption{Spacetime diagram illustrating the so called ``twin
        paradox'' in one spatial dimension and in units where $c=1$, so
        that light travels on $45^\circ$ lines. The numbers indicate the
        elapsed proper time along each trajectory, and the dots indicate
        clock ticks for the identical clocks carried by each twin.}
    \label{Fig:TwinPdx}
\end{figure}

The phenomenon of time dilation can be efficiently described in the
context of a thought experiment termed the ``twin paradox'' in special
relativity (despite the name, it is not strictly a paradox), which is
illustrated in a spacetime diagram in Fig. \ref{Fig:TwinPdx} (cf. Fig.
5-8 of \cite{TaylorWheeler:1992}). In this thought experiment, a pair of
twins, \textbf{TWIN I} and \textbf{TWIN II}, carry identical clocks that
count units of time---ticks. \textbf{TWIN I} remains stationary, and
\textbf{TWIN II} travels at relativistic speeds on a round trip journey.
When the twins reunite and compare their clocks, the clock of
\textbf{TWIN I} registers a greater number of ticks than \textbf{TWIN
II}.

This prediction of special relativity follows from the line element (cf.
Eq. \eqref{EqA:LineElement}):
\begin{equation} \label{Eq:LineElement}
    ds^2=-c^2dt^2+d\ubar{x}^2+d\ubar{y}^2+d\ubar{z}^2,
\end{equation}
where $(t,\ubar{x},\ubar{y},\ubar{z})$ are the spacetime coordinates in
the frame of \textbf{TWIN I} ($t$ corresponds to the clock time of
\textbf{TWIN I}), and $c$ is the speed of light. For theoretical
analyses, it is convenient to choose units where $c=1$ and to restrict
the motion to the $\ubar{x}$-direction, so that $d\ubar{y}=d\ubar{z}=0$;
this is done for the illustration in Fig. \ref{Fig:TwinPdx}. The line
element provides a measure of distances and times between
infinitesimally separated points (it is in some sense a generalization
of the Pythagorean theorem), and encapsulates the mathematical content
of special relativity. If $ds^2=0$, the points lie along a lightlike
trajectory (the path of a particle traveling at the speed of light), and
if $ds^2<0$, the points lie along a timelike trajectory, that is, the
path of a particle traveling slower than the speed of light. In the
latter, the elapsed proper (or clock) time between the infinitesimally
separated points is given by $c^2 d\tau^2 = -ds^2$. One can then measure
elapsed proper times along spacetime trajectories using the line
element. For a particle moving at a velocity $v=d\ubar{x}/dt$, one may
employ the trick of dividing by differentials to obtain the following
expression, which one might recognize as the Lorentz factor $\gamma$
(cf. Eq. \eqref{EqA:Four-Velocity-TimeComponent}):
\begin{equation} \label{Eq:LorentzFactorTD}
    \frac{dt}{d\tau}=\frac{1}{\sqrt{1-(v/c)^2}}=\gamma,
\end{equation}
It follows that as the velocity $v$ approaches the speed of light, a
given interval of proper time $\Delta\tau$ corresponds to a large
difference $\Delta t$ in coordinate time $t$. This can be seen by
examining the angles of the segments between clock ticks in the path of
\textbf{TWIN II} in Fig. \ref{Fig:TwinPdx}; the segments at shallow
angles closer to $45^\circ$ lines are further separated in the vertical
direction than endpoints on vertical segments. For a more in-depth
discussion of special relativity and the twin paradox, we refer the
reader to the excellent treatment in \cite{TaylorWheeler:1992}.

In general relativity, one works with a more general line element of the
form $ds^2=g_{\mu\nu} dx^\mu dx^\nu$, where $\mu,\nu \in \{0,1,2,3\}$
are coordinate indices (Einstein summation convention is employed here),
with the value $0$ indicating the time coordinate, and the coefficients
$g_{\mu\nu}$ are in general functions of the spacetime coordinates
$x^\mu$. The coefficients $g_{\mu\nu}$ form the components of the metric
tensor, which one obtains as solutions to a system of nonlinear coupled
partial differential equations known as the Einstein field equations
\cite{Carroll,Wald}. A well-known solution describing the spacetime
around a spherically symmetric massive object is the Schwarzschild
solution, which yields a line element of the form (cf. Eq.
\eqref{EqA:Schwarzschild}):
\begin{equation}\label{Eq:Schwarzschild}
    ds^2 = - \left[1-\frac{2 M}{r}\right] c^2 dt^2 
           + \frac{dr^2}{1-2 M/r} 
           + r^2 (d\theta^2+\sin^2\theta d\phi^2)
\end{equation}
in spherical polar coordinates $(t,r,\theta,\phi)$, where $M:=G
\bar{M}/c^2$ is the mass parameter (which has units of length, and a
value half the Schwarzschild radius), with $\bar{M}$ being the physical
mass and $G$ is Newton's constant. From the Jebsen--Birkhoff theorem
\cite{Birkhoff:1927relativity,Jebsen:1921general}, this line element is
valid outside of any spherically symmetric object absent the presence of
any other matter, and can describe the spacetime geometry outside
nonrotating stars and black holes \cite{Carroll,Wald}. In the case of
the latter, the event horizon is located at a radius $r=2M$. The motion
of free-falling massive particles can be obtained by finding geodesic
paths, that is, paths that maximize the elapsed proper time $\Delta \tau
= \int d\tau$ on each finite segment of the path.

Now consider a static observer corresponding to a path defined by
constant spatial coordinates $(r,\theta,\phi)$. Since,
$dr=d\theta=d\phi=0$ along such a path, one may readily obtain the time
dilation factor from Eq. \eqref{Eq:Schwarzschild}:
\begin{equation}\label{Eq:SchwarzschildTimeDilation}
   \frac{dt}{d\tau} = \frac{1}{\sqrt{1-2 M/r}}.
\end{equation}
At large values of $r$, ${dt}/{d\tau}\sim 1$, so that $t$ coincides with
the proper time of static observers far away from a massive object. In
the case of a black hole, the quantity ${dt}/{d\tau}$ becomes
arbitrarily large as the radial position $r$ approaches the event
horizon $2M$, so that a single tick of proper time $\tau$ corresponds to
an arbitrarily large interval of time $\Delta t$ as measured by a
faraway observer. This setup provides a relatively straightforward
example of gravitational time dilation.

However, such a setup is of limited utility. The path of a static
observer is nongeodesic, so that such observers are not in freefall; in
the case of the Earth, it corresponds to an observer sitting on the
surface. Such observers will therefore experience a local acceleration,
and one can show that the path at constant spatial coordinates
$(r,\theta,\phi)$ experiences a local outward radial acceleration with a
magnitude
\begin{equation}\label{Eq:SchwarzschildAcc}
    a_{st} = \frac{Mc^2}{r^2\sqrt{1-2 M/r}},
\end{equation}
the derivation of which may be found in Ch. 6 of \cite{Carroll} (cf. Eq.
6.22 therein). The acceleration $a_{st}$ diverges at the event horizon,
so that a static observer at the horizon will feel an infinite weight.
Any attempt to exploit time dilation in this manner is therefore unsafe
for biological matter with limited tolerance to a high acceleration
environment.

%=======================================================================
%-----------------------------------------------------------------------
%
%    Civilizations around supermassive black holes
%
%-----------------------------------------------------------------------
%=======================================================================

\section{Civilizations around supermassive black holes}
\label{sec:CivSMBH}

For biological matter, we seek a freefall trajectory which is locally
free of acceleration (tidal forces aside) in the sense of the
equivalence principle. Circular orbits around black holes are examples
of such trajectories. As discussed earlier, Kip Thorne considered stable
circular orbits around supermassive black holes with extreme spin.
However, such extreme spins are unlikely to occur in nature, so we
instead consider orbits around nonspinning black holes described by the
Schwarzschild line element in Eq. \eqref{Eq:Schwarzschild}. In this
case, there is an innermost stable circular orbit with a radius greater
than $r=6M$ (cf. Box 25.6 of \cite{MTW}). For these orbits, one can
crudely estimate the gravitational contribution to time dilation from
Eq. \eqref{Eq:SchwarzschildTimeDilation}, which is of order unity. It
follows that one cannot achieve large time dilation factors for stable
circular orbits around nonrotating black holes.

Instead of considering natural orbits around black holes with unnatural
parameters, it is perhaps more realistic to consider unnatural orbits
around black holes with natural parameters. That is, we consider
(artificially maintained) \emph{unstable} circular orbits, which can
have radii as small as $r=3M$ (the photon radius), and velocities
arbitrarily close to the speed of light. Such velocities can yield
arbitrarily large time dilation factors. For unstable circular orbits,
the formula for the time dilation factor is (cf. Eq.
\eqref{EqA:GammaFactorCirc}):
\begin{equation}\label{Eq:GammaFactorCirc}
    \Gamma:=\frac{dt}{d\tau}
           =\sqrt{\frac{r_\mathrm{c}}{r_\mathrm{c}-3M}},
\end{equation}
where $r_\mathrm{c}$ is the radius of the orbit. In the limit
$r_\mathrm{c}\rightarrow 3M$, $\Gamma$ can be arbitrarily large. In this
manner, one can find free falling trajectories that are arbitrarily time
dilated, compared to faraway observers.

While free falling trajectories have vanishing local acceleration,
objects of a finite size will still experience tidal forces. Tidal
forces can also become large in the limit $r_\mathrm{c}\rightarrow 3M$,
as slight differences in radii can lead to large variations in the time
dilation factor $\Gamma$. This is pertinent for biological organisms
which necessarily have finite extent, irrespective of their composition;
sufficiently large tidal forces can potentially disassemble biological
organisms above a characteristic size (which must be finite). 

The magnitude of tidal forces across a distance $\chi$ can be estimated
with the following formula for the relative tidal acceleration between
points separated by $\chi$ in directions transverse to the motion:
\begin{equation}\label{Eq:AccelComponentsCircDiv}
    a_{\mathrm{tidal}} \approx \frac{c^2 \chi}{9M (r_\mathrm{c}-3M)}
    =
    \frac{c^6 \left(\Gamma ^2-1\right) \chi }{27 G^2 \bar{M}^2}.
\end{equation}
This follows from the geodesic deviation equation
\cite{MTW,Pirani:1956tn,Synge:1934zza} applied to the aforementioned
unstable circular orbits in the Schwarzschild geometry; the detailed
derivation of this formula is provided in the appendix (cf. Eqs.
\eqref{EqA:AccelComponentsCirc} and \eqref{EqA:AccelComponentsCircDiv}).
We only consider tidal accelerations in the transverse directions, since
the tidal acceleration in the direction of motion (the $\phi$ direction)
is finite and bounded at the photon radius (in the limit $r_{\rm
c}\rightarrow3M$; see the discussion following Eq.
\eqref{EqA:AccelComponentsGen}). We seek conditions tolerable for
humans, and as such, we consider a distance of $\chi\approx2~\mathrm{m}$
corresponding to the size of humans \cite{ncd:2016century,Press:1980man}
and a relatively gentle acceleration of $1~\mathrm{m/s}^2$, or about one
tenth of the gravitational acceleration on the surface of the Earth (we
assume that large vessels can be engineered to tolerate much higher
accelerations).\footnote{One might be concerned about large tidal
stresses on the structure of a large vessel. At a radius $r$
corresponding to a tidal acceleration of $1~{\rm m/s}^2$ across $2~\rm
m$, the relative tidal acceleration across a vessel with a width/height
of $\sim 20~{\rm m}$ in the directions transverse to the direction of
motion increases to $10~\mathrm{m/s}^2$ (observe that Eq.
\ref{Eq:AccelComponentsCircDiv} is linear in $\chi$), which is roughly
the gravitational acceleration experienced on the surface of the Earth.
Centrifuges \cite{Hyde:1963manrated} and fighter jets
\cite{Agard:1995current} can tolerate stresses well in excess of this
limit (tens of $\rm g$). The extent of the vessel is far less
constrained in the direction of motion, as the tidal accelerations in
the direction of motion remain finite and bounded even at the photon
radius.}  Given these anthropocentric values, we can establish a
relationship between the time dilation factor $\Gamma$ and the black
hole mass $\bar{M}$.

\begin{figure}[t]
    \centering
    \includegraphics[width=0.50\textwidth]{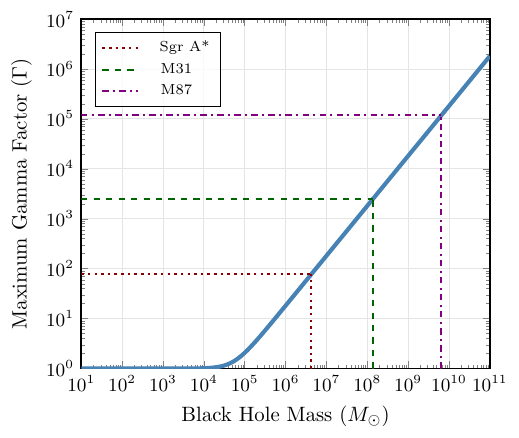}
    \caption{ Time dilation factor $\Gamma$ for an unstable circular
        orbit (near the photon radius) versus mass for a tidal
        acceleration $a_\mathrm{tidal}=1~\mathrm{m/s}^2$ across a
        distance of $\chi\sim2~\mathrm{m}$.}
    \label{Fig:GammavMass}
\end{figure}

In figure \ref{Fig:GammavMass}, we present a logarithmic plot of the
maximum time dilation factor $\Gamma$ (as a function of black hole mass
$\bar{M}$) that can be achieved for a limiting tidal acceleration of
$1~\mathrm{m/s}^2$ across a distance $\chi\approx2~\mathrm{m}$. We see
that stellar mass black holes, which are expected to top out at a few
hundred solar masses, provide very little time dilation (as
$\Gamma\sim1$). Any meaningful time dilation requires a supermassive
black hole.

Thus, our search for a significantly time-dilated and biologically
tolerable frame leads us to supermassive black holes, which includes Sgr
A*, the $4.3\times10^6~M_{\odot}$ supermassive black hole at the center
of the Milky Way galaxy \cite{GRAVITY:2023avo}. From Fig.
\ref{Fig:GammavMass}, we see that a time dilation factor of around $100$
is biologically tolerable around Sgr A*. We also see that the more
massive supermassive black holes at the center of the M31 (Andromeda)
and M87 galaxies permit higher values for $\Gamma$ by orders of
magnitude.

A time dilation factor of $100$ is relatively limited considering the
overall size of the Milky Way, with a diameter of $\sim
87,400~\mathrm{ly}$ \cite{Goodwin:1998}. On the other hand, the density
of stars near the galactic center is on the order of $0.28$ stars per
cubic light year, so there are still about one million stars within
$100~\mathrm{ly}$ of Sgr A*. A civilization at Sgr A* with a time
dilation factor of $100$ will have no shortage of destinations reachable
within a tolerable duration in the civilizational frame.

Observations of flares from Sgr A* indicate the presence of accreting
plasma \cite{GRAVITY:2023avo,Marin:2023qvc}. This would potentially
create a considerable drag force for an ultrarelativistic vessel on
which the inhabitants reside. In addition to slowing the vessels, such a
drag force would perturb the orbits from their unstable equilibria.

We can estimate the power required to counter\footnote{Though we refrain
from speculating on specific technologies, one might imagine that the
vessels are propelled by some form of beamed propulsion, requiring
autonomous infrastructure based elsewhere. We discuss this further at
the end of the article.} the effect of drag from accreting plasma by
assuming that the ions that collide with the vessel are accelerated to
the vessel's speed. Each colliding particle of mass $m$ will then
acquire a kinetic energy of $\sim \Gamma m c^2$. An ultrarelativistic
vessel of a given cross sectional area $A_\mathrm{cs}$ sweeps out a
volume of $\sim c A_\mathrm{cs}$ per unit time. The power required to
counter the drag force is then given by:
\begin{equation}\label{Eq:Pdrag}
    P_\mathrm{drag} \approx A_\mathrm{cs} \Gamma \rho c^3,
\end{equation}
where $\rho$ is the mass density of the accreting plasma. Assuming the
ions in the plasma are primarily Hydrogen (with a mass of $1.67\times
10^{-27}~\mathrm{kg}$), one can obtain the mass density $\rho$ from the
estimated density of gas particles around Sgr A*, which ranges from
$10^4$ to $10^{10}$ particles per cubic centimeter \cite{Hosseini_2020}.
If we further assume $\Gamma\sim 100$, and a cross sectional area of
about $A_\mathrm{cs}\sim1,000~\mathrm{m}^2$, we obtain a range for
$P_\mathrm{drag}$ of $4.51\times10^{13}~\mathrm{W}$ to
$4.51\times10^{19}~\mathrm{W}$.

For comparison, the current power capacity of human civilization is on
the order of $10^{13}~\mathrm{W}$ \cite{EI:2024statistical}, and the
defining power capacity for a Type I civilization on the Kardashev scale
\cite{Kardashev:1964} is typically given as $\sim10^{17}~\mathrm{W}$
\cite{Gray:2020ext,Lemarchand:1994det,Sagan:1973icar}. Remarkably, the
present global power capacity for human civilization is within an order
of magnitude required to counter the drag from accreting plasma in the
lower density limit of the scenario we have considered. Of course,
flaring activity near Sgr A* \cite{GRAVITY:2023avo,Marin:2023qvc}
suggests that the plasma can vary greatly in density over time, so that
a permanent civilization based at Sgr A* should at least have the
capacity to accommodate higher densities. We regard the higher power
capacity estimate $4.51\times10^{19}~\mathrm{W}$ (which exceeds the Type
I threshold by two orders of magnitude) to be a more realistic minimum
requirement for such a civilization.

\begin{figure}[htp]
    \centering
    \includegraphics[width=0.50\textwidth]{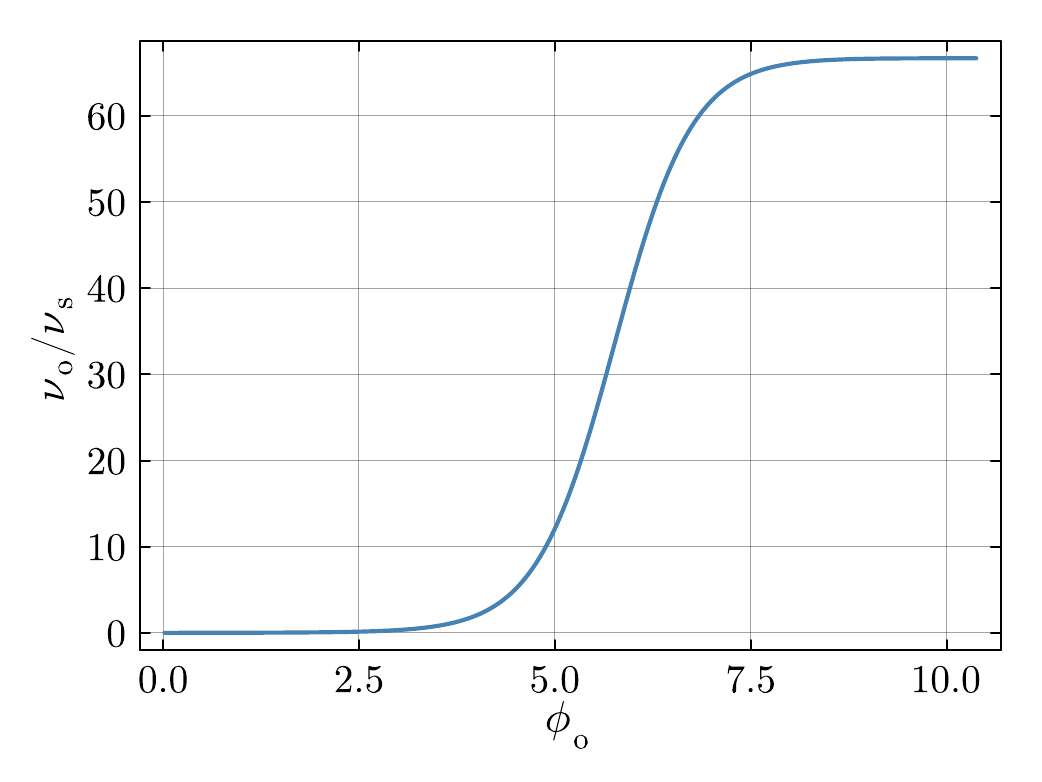}
    \caption{Frequency ratio $\nu_\mathrm{o}/\nu_\mathrm{s}$ versus
        angular position of a distant observer $\phi_\mathrm{o}$, given
        a source in a circular orbit with time dilation factor
        $\Gamma=10^4$ with $l>0$ emitting a signal isotropically with
        frequency $\nu_\mathrm{o}$. Since $l>0$, the source moves in the
        positive $\phi$ direction. As the source moves, the curve shifts
        to the right; an observer at a fixed angular position will see
        $\nu_\mathrm{o}/\nu_\mathrm{s}$ decrease as the source continues
        to move in the positive $\phi$ direction.}
    \label{Fig:FreqvAngle}
\end{figure}

We now consider some general features that one might expect for signals
from these orbits, as possible technosignatures from highly time-dilated
civilizations around supermassive black holes. In particular, we
consider a simple model for the source signal that is continuous,
monochromatic, and isotropic. The source is assumed to be in a circular
orbit of some fixed radius $r_\mathrm{c}$ near the photon radius of a
Schwarzschild black hole with mass parameter $M$. Given a source
frequency $\nu_\mathrm{s}$ and a distant observer in the orbital plane,
one may show that the observed frequency $\nu_\mathrm{o}$ and the
angular coordinate $\phi_\mathrm{o}$ of the observer can be expressed
parametrically as  [cf. Eqs. \eqref{EqA:PhotonFrequencyFactor},
\eqref{EqA:IntegralNull}]:
\begin{equation}\label{Eq:FrequencyAngleObs}
\begin{aligned}
    \nu_\mathrm{o}(\vartheta) 
    &= \frac{\nu_\mathrm{s} \sqrt{1-2 M/r_\mathrm{c}} 
    \left[\sqrt{r_\mathrm{c}(1-2 M/r_\mathrm{c})} 
            \mp \sqrt{M} \cos (\vartheta)
    \right]}{\sqrt{(r_\mathrm{c}-3 M)}}, \\
    \phi_\mathrm{o}(\vartheta)
    &=\int_{r_\mathrm{c}}^\infty
       \frac{B(\vartheta) ~ dr}{\sqrt{r(r^3-B(\vartheta)^2(r-2M))}},
\end{aligned}
\end{equation}
where $\vartheta$ is a parameter, and the following function is defined
(cf. Eq. \eqref{EqA:PhotonImpactParam}):
\begin{equation}\label{Eq:PhotonImpactParam}
    B(\vartheta)
    = 
    \frac{\pm \sqrt{M}-\cos (\vartheta)\sqrt{r_\mathrm{c}-2M} }
    {\sqrt{r_\mathrm{c}-2M}\mp\sqrt{M}\cos (\vartheta)}
    \frac{r_\mathrm{c}}{\sqrt{1-2 M/r_\mathrm{c}}}.
\end{equation}
The sign $\pm$ corresponds to direction of orbital rotation; in the case
where the orbital plane is the equatorial plane $\theta=\pi/2$ (in the
coordinates used in Eq. \eqref{Eq:Schwarzschild}), a positive sign
corresponds to an orbit moving in the direction of positive $\phi$. The
derivation of these formulas can be found in the appendix.

We plot Eq. \eqref{Eq:FrequencyAngleObs} relating $\nu_\mathrm{o}$ and
$\phi_\mathrm{o}$ in Fig. \ref{Fig:FreqvAngle}. This illustrates the
dependence of the observed frequency on the angular coordinate of the
observer (here, we assume both source and observer are in the equatorial
plane). The phase of $\phi_\mathrm{o}$ is tied to the angular coordinate
of the source, which for a circular orbit, depends linearly on time. As
the source moves, the curve will shift horizontally---in this way, one
can infer the time dependence of the observed frequency
$\nu_\mathrm{o}$. A source moving in the positive $\phi$ direction will
yield a curve that shifts to the right, so that an observer sitting at a
constant angle $\phi_\mathrm{o}$ will see a downward drift in the
frequency. Observe that the domain of the curve in Fig.
\ref{Fig:FreqvAngle} extends beyond $2\pi$. This implies that a signal
emitted at a given instant can reach the observer along multiple paths
(and at distinct observed frequencies). Physically, this corresponds to
the (possibly multiple) winding of signal trajectories around the black
hole before reaching the observer \cite{Darwin:1959gravity}.

Our analysis is, of course, limited to nonrotating Schwarzschild black
holes, and there are multiple studies indicating that supermassive black
holes like Sgr A* and M87 have a large spin parameter
\cite{Daly:2023axh}. The behavior of generic orbits for rotating black
holes is more involved, and while additional possibilities become
available, at least one of these has been considered before
\cite{Thorne:2014}, and a more comprehensive exploration of highly
time-dilated orbits with limited tidal acceleration in Kerr (or
Kerr-Newman) spacetimes is perhaps appropriate for a separate study.

%=======================================================================
%-----------------------------------------------------------------------
%
%    A galactic confederation from linear motion
%
%-----------------------------------------------------------------------
%=======================================================================

\section{A galactic confederation from linear motion}
\label{Sec:LinearMotion}

A more ambitious civilization may seek frames with a higher time
dilation factor. In particular, one might seek a time dilation factor of
$\sim 10^4$ or higher so that trips spanning a significant fraction of
the Milky Way diameter may be completed on a time scale of decades in
the civilizational frame. In this section, we describe how such a
civilization may be constructed.

\begin{figure}[t]
    \centering
    \includegraphics[width=0.50\textwidth]{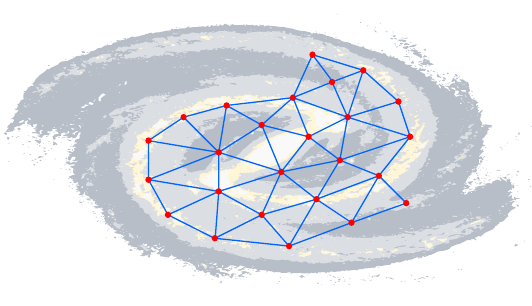}
    \caption{ An illustration of a civilization based around a network
        of linear paths and nodes. Vessels accelerate and decelerate
        along each path, such that multiple vessels simultaneously
        arrive at each node when the civilizational frame coincides with
        that of the Milky Way. }
    \label{Fig:CivNet}
\end{figure}

\begin{figure}[htpb]
    \centering
    \includegraphics[width=0.50\textwidth]{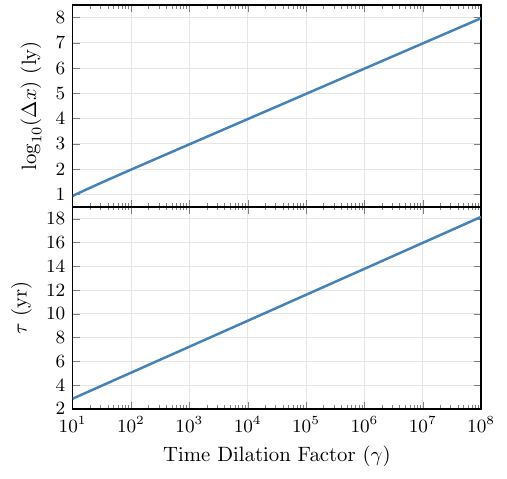}
    \caption{ Plots of distance and proper time required to accelerate a
        vessel to a peak time dilation factor $\gamma_\mathrm{p}$,
        assuming a uniform proper acceleration of $10~\mathrm{m/s}^2$.
        The top subplot corresponds to Eq. \eqref{Eq:PropTravelTime} and
        the bottom corresponds to Eq. \eqref{Eq:DistanceTraveled}.}
    \label{Fig:LinMot}
\end{figure}

Recall Sagan's \cite{Sagan:1963} observation that a vessel accelerating
at $\sim 10~\mathrm{m/s}^2$ can reach any point in the Milky Way galaxy
within the lifetime of a human crew due to special relativistic time
dilation. We consider Sagan's special relativistic model for linearly
accelerating motion, in which a vessel accelerates at $\sim
10~\mathrm{m/s}^2$ until reaching a desired velocity at the midpoint of
its journey before decelerating in the same manner back to the galactic
frame. We propose that a civilization can be constructed from a network
of vessels (as illustrated in Fig. \ref{Fig:CivNet}) undergoing such a
motion between a set of nodes, synchronized so that several vessels
simultaneously meet at each node when they decelerate to the galactic
frame, and that each vessel experiences the same amount of proper time
during each trip. The individuals comprising the civilization reside on
these vessels, and an individual may choose to either remain within a
vessel or transfer to another vessel when the vessels meet. In this
section, we will demonstrate that a traveler can in this manner visit
multiple nodes on a voyage spanning the diameter of the Milky Way within
a single human lifetime.

In the remainder of this section, we establish the basic parameters of
the motion and estimate the power requirements for such a civilization.
The basic formulas for the motion of a uniformly accelerated vessel in
special relativity are given in the appendix. From Eq.
\eqref{EqA:Hyperbolic} in the appendix, the motion of a uniformly
accelerated vessel in the direction of the $x$ coordinate can be
described parametrically as:
\begin{equation} \label{Eq:Hyperbolic}
    X^\mu(\tau) = \left(\frac{c^2 \sinh(a \tau/c)}{a},
                    \frac{c^2 (\cosh(a \tau/c)-1)}{a},0,0\right),
\end{equation}
where $X^\mu$ describes the spacetime coordinates $(t,x,y,z)$ of the
vessel (the index $\mu$ runs over the range $\mu \in \{0,1,2,3\}$, with
$X^0=t$, $X^1=x$, and so on). The quantity $\tau$ represents the proper
time as measured by observers on the vessel, and $a=\sqrt{|\eta_{\mu\nu}
a^\mu a^\nu|}$ (here and throughout we employ Einstein summation
convention) is the magnitude of the proper acceleration
$a^\mu=d^2X^\mu/d\tau^2$ experienced by the observer. It is not too
difficult to verify that $a$ is constant. One may take the derivative of
Eq. \eqref{Eq:Hyperbolic} to obtain the four-velocity:
\begin{equation} \label{Eq:Hyperbolic4vel}
    u^\mu(\tau) := \frac{dX^\mu}{d\tau} = \left(c \cosh(a \tau/c),
                    c \sinh(a \tau/c),0,0\right).
\end{equation}
The special relativistic time dilation factor corresponds to the
component $u^0$ which is given by the Lorentz factor [cf.
Eq. \eqref{EqA:Four-Velocity-TimeComponent}]:
\begin{equation} \label{Eq:LorentzFactor}
    \gamma=\frac{dt}{d\tau}=c \cosh(a\tau/c)=\frac{1}{\sqrt{1-(v/c)^2}},
\end{equation}
where $v$ is the magnitude of the spatial velocity
$v^i=dX^i/dt=u^i/\gamma$ (with $i\in\{1,2,3\}$ being a spatial index).
Equation \eqref{Eq:LorentzFactor} can in principle be used to compute
the proper travel time $\tau_\mathrm{p}$ required to reach a peak
Lorentz factor $\gamma_\mathrm{p}$:
\begin{equation} \label{Eq:PropTravelTime}
    \tau_\mathrm{p}=\frac{c}{a} \cosh^{-1}(\gamma_\mathrm{p}).
\end{equation}
The above expression may be used with Eq. \eqref{Eq:Hyperbolic4vel} to
calculate the distance traveled:
\begin{equation} \label{Eq:DistanceTraveled}
    \Delta x_\mathrm{p}=\frac{c^2}{a} (\gamma_\mathrm{p}-1).
\end{equation}
We plot Eqs. \eqref{Eq:PropTravelTime} and \eqref{Eq:DistanceTraveled}
in Fig. \ref{Fig:LinMot}, assuming a proper acceleration of
$10~\mathrm{m/s}^2$. For this acceleration, a time dilation factor of
$\gamma_\mathrm{p}\sim 10^4$ is achieved in a proper travel time of
$\tau_\mathrm{p}\sim 10~\mathrm{yr}$, and a travel distance of $\Delta
x_\mathrm{p} \sim 10^4~\mathrm{ly}$. 

The proper time required to accelerate to a peak time dilation factor of
$\gamma_\mathrm{p}\sim 10^4$ and decelerate back to the galactic frame
is $20~\mathrm{yr}$, and the distance covered in this time is $\sim
20,000~\mathrm{ly}$, between a fourth and fifth of the diameter of the
Milky Way \cite{Goodwin:1998}. In the context of the scenario described
earlier and illustrated in Fig. \ref{Fig:CivNet}, an individual
traveling on linearly accelerating vessels (accelerating at
$10~\mathrm{m/s}^2$ and with a peak factor $\gamma_\mathrm{p}\sim 10^4$)
can travel a distance equivalent to the diameter of the Milky Way in a
century as measured in the civilizational frame.

We now consider the power requirements for accelerating a vessel to a
time dilation factor $\gamma_\mathrm{p}\sim 10^4$. From Eqs.
\eqref{EqA:Power} and \eqref{EqA:Four-Acceleration} and the derivative
of \eqref{Eq:Hyperbolic4vel}, one can obtain the following expression
for the power required to accelerate a vessel of mass $m$ on the
spacetime trajectory given in Eq. \eqref{Eq:Hyperbolic4vel}:
\begin{equation} \label{Eq:PowerLinAcc}
    P_\mathrm{acc}=cma\tanh(a\tau/c)=cma\sqrt{1-\gamma^{-2}}.
\end{equation}
Assuming a vessel of mass $10^{13}~\mathrm{kg}$ (roughly that of a
kilometer-size asteroid), an acceleration of $10~\mathrm{m/s}^2$ and a
peak time dilation factor $\gamma_\mathrm{p}\sim 10^4$, the peak power
required to maintain the vessel in uniform acceleration is on the order
of $10^{22}~\mathrm{W}$.

Additionally, one must consider the power required to counteract the
drag force from the interstellar medium, which is given by $
P_\mathrm{drag}= A_\mathrm{cs} \rho c^2 \gamma v$, where $A_\mathrm{cs}$
is the cross-sectional area of the vessel, and $\rho$ is the density of
the interstellar medium. This expression can be simplified slightly
since for $\gamma_\mathrm{p}\gg1$, $v_\mathrm{p}\sim c$, so that for
large time dilation factors, one has:
\begin{equation} \label{Eq:PowerDrag}
    P_\mathrm{drag} \approx A_\mathrm{cs} \rho \gamma c^3,
\end{equation}
Now the interstellar medium can have molecular densities up to $\sim
10^{12}~\mathrm{m}^{-3}$ in molecular clouds \cite{Ferriere:2001rg},
which we assume to be primarily composed of hydrogen molecules. Given a
cylindrical vessel $100~\mathrm{m}$ in diameter and a gamma factor
$\gamma_\mathrm{p}\sim 10^4$, we estimate that the power
$P_\mathrm{drag}$ required to counteract drag from the interstellar
medium is on the order of $10^{18}~\mathrm{W}$, four orders of magnitude
less than that the peak acceleration power (and well below the
luminosity of a brown dwarf).

From these power estimates, we find that a single Type II civilization
has the power capacity ($L_{\odot}=3.83\times10^{26}~W$) to
accelerate\footnote{As in the preceding section, we imagine that these
vessels are propelled by some form of beamed propulsion from autonomous
infrastructure based at stars along the path.} tens of thousands of
vessels on linear trajectories satisfying the criteria we employed in
the preceding analysis. From power requirements alone, a single Type II
civilization should have the capacity to establish a rather substantial
galactic confederation consisting of tens of thousands of vessels
traveling on linear trajectories between nodes spaced apart by a few
tens of thousands of light years (as caricatured in Fig.
\ref{Fig:CivNet}).  

%=======================================================================
%-----------------------------------------------------------------------
%
%    A ring of black holes
%
%-----------------------------------------------------------------------
%=======================================================================

\section{A ring of black holes} \label{Sec:RingBH} 

\begin{figure}[htpb]
    \centering
    \includegraphics[width=0.50\textwidth]{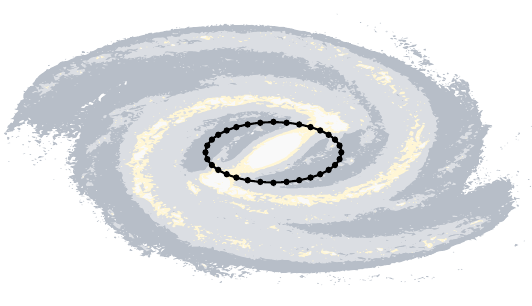}
    \caption{ Illustration of a ring of black holes concentric with the
        center of the Milky Way. }
    \label{Fig:CivRing}
\end{figure}

A civilization with greater ambitions and a higher power capacity may
seek to establish a more permanent residence. One option is to construct
a ring of black holes (see Fig. \ref{Fig:CivRing}) to form a trajectory
that scatters off each black hole so that it forms a closed spatial path
within the Milky way. The primary advantage of such an approach is that
it avoids the need to constantly accelerate and decelerate all of the
vessels on which the civilization resides; one only needs to counteract
the drag force from the interstellar medium and to maintain the
positions of the black holes in the ring. Moreover, if the black holes
have a significant amount of spin, they can serve as enormous sources of
energy via the Penrose process \cite{Penrose:1971uk}. Following the
procedure in Sec. 12.4 of \cite{Wald}, one can show that the rotational
energy for a \emph{single} black hole with a spin parameter close to the
maximal limit and a mass of $14~M_\odot$ (which are average values for
stellar mass black holes \cite{KAGRA:2021duu,Olejak:2019pln}) would be
enough to power a Type II civilization thousands of times the age of the
universe.\footnote{This should not be surprising, as stars lose a
relatively small fraction of their mass-energy over their lifetimes, and
for a star rotating close to the maximal limit, up to $29\%$ of its
mass-energy can in principle be extracted via the Penrose process.} Once
such a structure is constructed, it may serve as the seat for a
centralized authority for an existing network of linearly accelerating
vessels. A ring of black holes may therefore provide an ideal foundation
for establishing a more permanent and robust civilization---a galactic
empire. Another potential motivation for constructing such a structure
arises from scientific curiosity; a galaxy-scale ring of black holes can
be used to maintain elementary particles on a circular trajectory---one
might imagine that such a structure may facilitate the construction of a
galaxy scale particle accelerator needed to directly probe Planck scale
physics \cite{Lacki:2015lda}.

We begin by considering the problem of scattering from a single,
non-rotating black hole of mass $\bar{M}$, the mathematical details of
which may be found in the appendix. From Eq.
\eqref{EqA:DeflectionAngle}, the angle of deflection, that is, the angle
by which the particle deviates from a straight line, is given by: 
\begin{equation}\label{Eq:DeflectionAngleGeneral}
    \Delta\psi
    \approx
    \frac{2G\bar{M}(2 e^2-c^2)}{r_\mathrm{c}(e^2-c^2)c^2},
\end{equation}
which is valid as long as the point of closest approach $r_\mathrm{c}$
is much larger than the Schwarzschild radius
$r_\mathrm{s}=2G\bar{M}/c^2$. The quantity $e$ is a conserved quantity
for free falling motion, which is related to the time dilation (Lorentz)
factor far from the black hole \footnote{Far from the black hole,
spacetime is approximately flat, so that special relativistic
considerations apply.}, that is $e\approx\gamma c^2$. Now for large
values of $\gamma$, the deflection angle coincides with the well-known
expression for light deflection:
\begin{equation}\label{Eq:DeflectionAngleLight}
    \Delta\psi
    \sim
    \frac{4G\bar{M}}{c^2 r_\mathrm{c}}.
\end{equation}
At the point of closest approach, the magnitude of the tidal
acceleration is given by [cf. Eqs. \eqref{EqA:AccelComponentsSc} and
\eqref{EqA:AccelComponentsScDiv}]:
\begin{equation}\label{Eq:ScatteringTidalAccel}
    a_{\mathrm{sc}} 
    \approx 
    \frac{3 G \bar{M} \gamma^2 \chi}{r_\mathrm{c}^3},
\end{equation}
where $\chi$ is the size of the body experiencing the tidal forces.
Equations \eqref{Eq:DeflectionAngleLight} and
\eqref{Eq:ScatteringTidalAccel} may be combined to obtain the following
relation:
\begin{equation}\label{Eq:DeflectionAngleAccel}
    \Delta\psi
    \sim
    \left(
        \frac{64 a_{\mathrm{sc}} G^2 \bar{M}^2}{3\gamma^2\chi c^4}
    \right)^{1/3}.
\end{equation}
One can see from the above expression that limits on the tidal
acceleration $a_{\mathrm{sc}}$ place strong limits on the maximum
deflection angle allowed for each scattering event.

Given a maximum allowable deflection angle $\Delta\psi$, we now consider
the number of black holes needed to deflect the trajectory of a vessel
into a closed path. In particular, we require that the sum of the
deflection angles $\Delta\psi_I$ is equal to $2\pi$:
\begin{equation}\label{Eq:TwoPi}
    \sum_{I=1}^N\Delta\psi_I = 2\pi,
\end{equation}
where $N$ is the number of black holes in the ring of black holes. To
simplify the analysis, we assume the black holes all have the same mass,
so that $N\Delta\psi_I=2\pi$; one may readily solve for $N$ to obtain:
\begin{equation}\label{Eq:NumberofBHs}
    N =
    2\pi \left(
            \frac{3\chi\gamma^2 c^4}{64 a_{\mathrm{sc}} G^2 \bar{M}^2}
         \right)^{1/3}.
\end{equation}
Assuming a time dilation factor of $\gamma=10^4$, an acceleration of
$a_{\mathrm{sc}} \sim 1~\mathrm{m/s}^2$ (one tenth of that of the
acceleration on Earth), a body size of $\chi\sim 2~\mathrm{m}$, and an
average black hole mass of $\sim 14~M_\odot$ \cite{Olejak:2019pln}, one
finds that at least $N\approx 7.9\times 10^5$ black holes are required
to deflect the trajectory into a closed spatial path. In its own frame,
the vessel encounters a black hole with a frequency of:
\begin{equation}\label{Eq:BHencounterFreq}
    f = \frac{\gamma c N}{2\pi r_\mathrm{ring}},
\end{equation}
where $r_\mathrm{ring}$ is the radius of the ring of black holes. For a
radius of $10,000~\mathrm{ly}$, this corresponds to a frequency of
$f\approx 4 \times 10^{-3}~\mathrm{Hz}$. This corresponds to a black
hole encounter roughly once every four minutes in the frame of the
vessel, and during each encounter, the occupants of the vessel
experience a momentary impulse of acceleration $a_{\mathrm{sc}} \sim
1~\mathrm{m/s}^2$ (as before, we assume a large vessel can be engineered
to tolerate much higher accelerations); one might expect such impulses
to be tolerable for terrestrial lifeforms on long timescales.

Now for a circular ring of $N\approx 7.9\times 10^5$ black holes with a
circumference $\pi\times10^4~\mathrm{ly}$, the black holes must be
spaced apart by a distance of $\sim 0.05~\mathrm{ly}$. The density of
black holes in the Milky Way is too low to naturally form a closed
trajectory. In the neighborhood of the sun, which is situated at a
radius $r_\odot=27,000~\mathrm{ly}$ from the galactic center, the
density of stellar black holes has been estimated to be
$\sim10^{-5}~\mathrm{ly}^{-3}$ \cite{Chisholm:2002ba,Shapiro:1983du}. It
is perhaps appropriate to assume that the density profile of stellar
mass black holes is proportional to the stellar disk density, which
takes the exponential form \cite{Gilmore:1983bv,McMillan:2016jtx}:
\begin{equation}\label{Eq:BaryonicMassDensity}
    \rho_\mathrm{d} \propto \exp(-r/r_\mathrm{d}),
\end{equation}
with a scale length of $r_\mathrm{d}\sim 8.5\times10^3~\mathrm{ly}$
\cite{McMillan:2016jtx}. This model indicates that at a radius of
$r\sim10,000~\mathrm{ly}$, the density of black holes increases by a
factor of $10$ to about $n_\mathrm{BH}\sim 10^{-4}~\mathrm{ly}^{-3}$.
This confirms that the stellar mass black hole density is too low to
naturally form a closed trajectory in the Milky Way. 

A civilization must therefore move black holes to form a ring that can
deflect the trajectory of a vessel into a closed trajectory. We consider
a ring of radius $r_\mathrm{ring}$ and a toroidal region around it
extending to a (minor) radius of $r_\mathrm{min}$. The number of black
holes contained in this region is:
\begin{equation}\label{Eq:NumberToroid}
    N \approx 2 \pi^2 r_\mathrm{ring} r_\mathrm{min}^2 n_\mathrm{BH}.
\end{equation}
For a ring of radius $r_\mathrm{ring}\sim10,000~\mathrm{ly}$, a density
$n_\mathrm{BH}\sim10^{-4}~\mathrm{ly}^{-3}$, and a number $N\approx
7.9\times 10^5$, one finds a minor radius $r_\mathrm{min}\sim
200~\mathrm{ly}$. To form a ring of black holes with a radius
$r_\mathrm{ring}\sim10,000~\mathrm{ly}$ (and concentric with the Milky
way), one must move $N\approx 7.9\times 10^5$ black holes a distance of
up to two hundred light years.

\begin{figure}[t]
    \centering
    \includegraphics[width=0.50\textwidth]{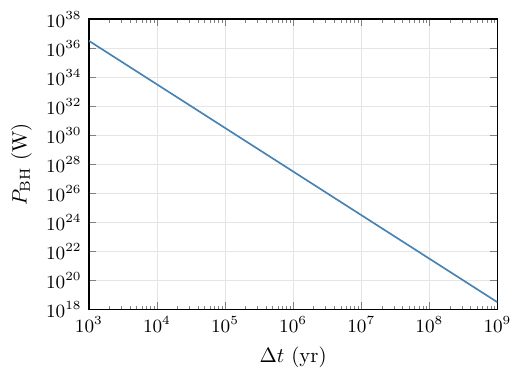}
    \caption{Logarithmic plot of Eq. \eqref{Eq:PowerBH} for the peak
    power $P_\mathrm{BH}$ required to move a black hole of mass
    $14~M_\odot$ a distance of $200~\mathrm{ly}$ in a time $\Delta t$.}
    \label{Fig:PowerBH}
\end{figure}

We now estimate the power requirements for moving the stellar mass black
holes. One may employ the model for motion in Sec.
\ref{Sec:LinearMotion}, setting $\Delta x=100~\mathrm{ly}$ in Eq.
\eqref{Eq:DistanceTraveled} to obtain an expression relating
$\gamma_\mathrm{p}$ and the acceleration $a$. The time required to
transport a black hole a distance $2\Delta x$ may be obtained from Eq.
\eqref{Eq:Hyperbolic}, and the identity $\sinh(x)^2=\cosh(x)^2-1$; we
obtain the expression:
\begin{equation}\label{Eq:Time}
    \Delta t 
    =\frac{2c}{a}\sqrt{\gamma^2_\mathrm{p}-1}
    =2\frac{\sqrt{\Delta x(\Delta x+2c^2/a)}}{c}.
\end{equation}
One may use the above expression to express $a$ and
$\gamma_\mathrm{p}$ in terms of $\Delta t$ and $\Delta x$ to obtain from
Eq. \eqref{Eq:PowerLinAcc} an expression for the peak power
$P_\mathrm{BH}$ required to transport a black hole of mass $\bar{M}$ a
distance of $2\Delta x$:
\begin{equation}\label{Eq:PowerBH}
    P_\mathrm{BH} 
    =\frac{4 c^4 \Delta t \Delta x^2\bar{M}}{c^4 \Delta t^4-\Delta x^4}.
\end{equation}
For a black hole of mass $\bar{M}=14 M_\odot$ and a distance $\Delta
x\sim 100~\mathrm{ly}$, we plot the required peak power $P_\mathrm{BH}$
as a function of transport time $\Delta t$ in Fig. \ref{Fig:PowerBH}. 

We see from Fig. \ref{Fig:PowerBH} that transporting a \emph{single}
$\bar{M}=14 M_\odot$ a distance of $200~\mathrm{ly}$ in two million
years requires a peak power of $P_\mathrm{BH}\sim 10^{26}~\mathrm{W}$
(on the order of the solar luminosity $L_{\odot}=3.83\times10^{26}~W$).
Moving $N\approx 6.2\times 10^5$ black holes of the same mass over the
same distance requires a total power capacity approaching
$10^{32}~\mathrm{W}$, which is still five orders of magnitude below the
power capacity of a Type III civilization ($\sim 10^{37}~\mathrm{W}$).
While a time scale of two million years may seem rather excessive for an
infrastructure project, there are two mitigating factors. The first is
that, as discussed earlier, stellar mass black holes contain enormous
amounts of rotational energy, and the Penrose process (or its
superradiant counterparts
\cite{Brito:2015oca,Cardoso:2004nk,Press:1972zz}) may facilitate a rapid
increase in a civilization's power capacity. The second is that the time
scale estimate is computed in the galactic rest frame. For a
civilization already in a frame that is time dilated by a factor of
$\gamma=10^4$, the time scale of such a project is reduced to about two
hundred years. This is lower than the time scales for the longest human
construction projects: the Great Wall of China was constructed over two
millennia \cite{Lovell:2007great}, and some European cathedrals, the
Milan Cathedral for instance, have required more than half a millennium
of continuous work to complete \cite{Sebregondi:2016}.

%=======================================================================
%-----------------------------------------------------------------------
%
%    Fermi paradox and conflict hazards
%
%-----------------------------------------------------------------------
%=======================================================================

\section{Fermi paradox and conflict hazards} \label{sec:Fermi}

Our considerations have so far been anthropocentric, but it is natural
to extend these considerations to extraterrestrial civilizations. A
discussion of the Fermi paradox is therefore unavoidable---we begin this
section with a brief review.

The Fermi paradox is often attributed to Enrico Fermi
\cite{Lytkin:1995,Jones:1985}, though it has been considered decades
before by Tsiolkovsky \cite{Tsiolkovsky:1933,Gray:2015} (who himself
attributes these ideas to other unnamed individuals). In particular, the
Fermi paradox refers to order of magnitude estimates of the number of
extraterrestrial civilizations based on the Drake equation
\cite{Drake:1961}, which often results in the conclusion that the Earth
must have been visited multiple times by extraterrestrial civilizations.
The lack of evidence for such visits typically leads one to conclude
either that intelligent life in the galaxy is exceedingly rare
\cite{Hart:1975,Tipler:1980,Tipler:1981} or that extraterrestrial life
is subject to a ``Great Filter'' that halts their progress in exploring
the Milky Way \cite{Hanson:1998}. Of course, many of the potential
explanations have been proposed, and the reader may find a somewhat
comprehensive compilation in \cite{Webb:2015}.

Here, we consider the proposition that extraterrestrial civilizations
seek time dilated frames (as in \cite{Reiss:2020,Li:2022}) as they move
up the Kardashev scale, and the implications of this proposition for
Fermi paradox. An extraterrestrial civilization with a capacity to send
expeditions to the Earth with a frequency $f$ and occupying a frame with
a relative time dilation factor of $\gamma$ would in the terrestrial
frame be expected to send expeditions at a reduced frequency of
$f_\oplus=f/\gamma$. In particular, a civilization in a frame with
$\gamma=10^4$ and a capacity to send trips to Earth annually in its own
frame would to terrestrial observers only be capable of sending an
expedition once every $10,000$ years (as measured on Earth), a timescale
comparable to the age of human civilization.

Of course, one might expect sub-Type II civilizations to migrate to Sgr
A*, and occupy frames with lower time dilation factors ($\gamma
\sim 10^2$ for human-like lifeforms). However, such civilizations would,
as argued earlier, have little interest in exploring regions far from
Sgr A* until they develop the capacity to access frames with higher time
dilation factors. Moreover, a large concentration of civilizations
surrounding Sgr A* might render the orbital lanes around Sgr A* a scare
resource---the potential for conflict increases with the number of
civilizations occupying orbits around Sgr A*. Civilizations based around
Sgr A* may therefore be reluctant to advertise their presence to other
civilizations in the Milky Way.

Now consider a scenario in which there are many galaxy-spanning
civilizations occupying high time dilation factors. A large number of
such civilizations may compensate for the reduced frequency of visits
due to the effect of time dilation. However, there is another
consideration that in our estimation has a much greater bearing on the
Fermi paradox. A civilization consisting of ultrarelativistic vessels
have a potentially severe vulnerability in that once their trajectories
are known, they may be relatively easy to destroy. For instance, a
$2~\mathrm{kg}$ object at rest in the galactic frame has a rest energy
equivalent to $43$ megatons of TNT, roughly that of the largest
thermonuclear device tested (the Soviet ``Tsar bomba''
\cite{Wellerstein:2021unearthly}), and in the frame of a vessel with a
time dilation factor of $10^4$ (relative to the galactic frame), the
same $2~\mathrm{kg}$ object will have a relative kinetic energy roughly
$10^4$ times its rest energy. To a vessel traveling at $\gamma=10^4$, a
$100~\mathrm{kg}$ object at rest in the galactic frame will have a
relative kinetic energy of $\sim 10^{23}~\mathrm{J}$, roughly the same
order of magnitude as the kinetic energy of the Chicxulub impact
\cite{Richards:2015tri}.

The maneuverability of an ultrarelativistic vessel is limited by
acceleration limits; one might worry that a hostile civilization with
knowledge of the vessel's trajectory can with relatively little energy
expenditure place a number of sufficiently massive and hard to detect
objects in the pathway of the vessel. In the case of a vessel in a frame
with a time dilation factor of $\gamma \sim 10^4$, the turning radius is
extraordinarily large due to the acceleration constraints. Circular
motion in special relativity may be described by the following
parametric expression (cf. \eqref{EqA:CircularSR}):
\begin{equation} \label{Eq:CircularSR}
    X_\mathrm{\circ}^\mu(\tau) 
        = \left( 
                c \gamma_\mathrm{\circ} \tau,
                R_\mathrm{\circ} \cos(\omega_\mathrm{\circ}\tau),
                R_\mathrm{\circ} \sin(\omega_\mathrm{\circ}\tau),
                0\right),
\end{equation}
where $\gamma_\mathrm{\circ}$ is the time dilation factor, $\tau$ is the
proper time as measured in the vessel's frame. The turning radius
$R_\mathrm{\circ}$ and frequency $\omega_\mathrm{\circ}$ satisfy the
following formulas (cf. \eqref{EqA:CircularSRrels}):
\begin{equation} \label{Eq:CircularSRrels}
    R_\mathrm{\circ} 
    = \frac{c^2 (\gamma_\mathrm{\circ}^2-1)}{a_\mathrm{\circ}},
    \qquad
    \omega_\mathrm{\circ} 
    = \frac{a_\mathrm{\circ}}{c \sqrt{\gamma_\mathrm{\circ}^2-1}},
\end{equation}
where $a_\mathrm{\circ}$ is the acceleration experienced by the vessel.
For $a_\mathrm{\circ}=10~\mathrm{m/s}^2$ and
$\gamma_\mathrm{\circ}=10^4$, one has an extraordinarily large turning
radius of $R_\mathrm{\circ}\approx9.5 \times10^7~\mathrm{ly}$, and a low
frequency of $\omega_\mathrm{\circ}\approx3.3\times10^{-12}~s$. For
$\tau \ll 1/\omega_\mathrm{\circ}$, the transverse motion may be
approximated as:
\begin{equation} \label{Eq:TransverseMotion}
    \Delta x_{\perp} 
        =  R_\mathrm{\circ} (1-\cos(\omega_\mathrm{\circ}\tau)) 
        \approx 
        \tfrac{1}{2}R_\mathrm{\circ} \omega_\mathrm{\circ}^2\tau^2 
        = \tfrac{1}{2} a_\mathrm{\circ} \tau^2 
        = \frac{a_\mathrm{\circ} t^2}{2 \gamma_\mathrm{\circ}^2},
\end{equation}
where $t$ is the time of an observer in the galactic frame. In the
galactic frame, a vessel limited to an acceleration of
$a_\mathrm{\circ}=10~\mathrm{m/s}^2$ can move transversely by less than
$50,000~\mathrm{km}$ in a single year (roughly twice the orbital radius
of the satellites comprising the GPS system), which translates to less
than an hour in the vessel's frame. Of course, any signals from the
vessel will arrive only $0.15~\mathrm{s}$ before the vessel itself, so a
hostile civilization intending to destroy the vessel must be able to
precisely predict the trajectory of the vessel. The vessel can therefore
mitigate threats by following an unpredictable trajectory.

While such vulnerabilities might be mitigated, potentially hostile
adversaries can still pose a threat by scattering masses in the pathway
of a vessel. Exacerbating the threat is the fact that potential
adversaries can emerge on a relatively short timescale. Human
civilization progressed from a hunter--gatherer society to one with the
capacity to place objects into interstellar space (specifically the
$800~\mathrm{kg}$ Voyager probes \cite{Voyager_Mission:2025}) within a
period of approximately $10,000$ years \cite{Harari:2014sapiens}; a
vessel traveling roughly towards the Earth at $\gamma \sim 10^4$ would
see the entire history of our present civilization in less than a year.
The rapid emergence of such threats would compel a civilization
occupying a high $\gamma$ frame to remain undetectable to other
civilizations. The strategy of hiding from other potentially hostile
civilizations has been explored before
\cite{Yu:2015JBIS,Webb:2015,Brin:1983}, and the scenario in which
extraterrestrial civilizations employ such a strategy has been termed
the ``Dark Forest'' resolution to the Fermi paradox
\cite{Yu:2015JBIS,Hibberd:2025}. An alternative (and perhaps more
disturbing) strategy is that of the Berserker hypothesis
\cite{Webb:2015}, in which a civilization sends out probes to
preemptively and systematically neutralize potentially hostile
civilizations (which in turn accounts for the lack of evidence for
extraterrestrial civilizations). The vulnerability of redshifted
civilizations to the rapid emergence of adversarial civilizations
provides a strong motivation for the Dark Forest or Berserker
hypotheses.

%=======================================================================
%-----------------------------------------------------------------------
%
%		SUMMARY AND DISCUSSION
%
%-----------------------------------------------------------------------
%=======================================================================
\section{Summary and discussion}\label{sec:conc}

In this article, we have shown that an advanced civilization can in
principle exploit time dilation to facilitate its exploration and
expansion up to a size approaching the diameter of the Milky Way within
the constraints of classical general relativity and a limited biological
tolerance to high acceleration environments. Moreover, we have
identified several ways in which a civilization can migrate to
time-dilated frames. As these are ordered according to the power
capacity of the civilization, one might regard these as a crude
description for the evolution of such a civilization as it advances on
the Kardashev scale.

We have described how a sub-Type II civilization can maintain an orbit
near the photon radius of Sgr A* with a time dilation factor of up to
$\gamma\sim100$ before the tidal acceleration becomes intolerable for
terrestrial biology. Such a time dilation factor enables round trips to
stars a few hundred light years from Sgr A* in a few years time as
measured in the civilizational frame. Given the stellar density of the
galactic center, the number of stars within reach in the aforementioned
timescale is on the order of a million, making the migration to Sgr A* a
somewhat compelling prospect, at least for civilizations that emerge
within a few thousand light years of the galactic center. For an annual
growth rate of 1\% in power capacity, human civilization will reach the
Type II threshold in about $3,000~\mathrm{yr}$, an order of magnitude
less than the time required to travel to Sgr A*, so civilizations with a
similar growth rate and located far from the galactic center may be less
inclined to migrate to Sgr A*.\footnote{On the other hand, there is some
recent evidence from pulsar data for a dark, massive object of mass
$\sim2.45 \times 10^7~M_\odot$ less than $3,000~\mathrm{ly}$ from the
Sun \cite{Chakrabarti:2025a}, which could be a runaway supermassive
black hole, like those seen in other galaxies \cite{vanDokkum:2023wed}.
If this is indeed the case, it would be a rather appealing destination
for human civilization, yielding a time dilation factor of several
hundred.}

A type II civilization can in principle establish a galaxy spanning
confederation. We employ Carl Sagan's model for the linear motion of an
ultrarelativistic spacecraft \cite{Sagan:1963}, in which a vessel
undergoes a period of uniform acceleration to a high time dilation
factor, the decelerates back to the galactic rest frame. A
straightforward analysis of the power requirements indicates that a Type
II civilization can in principle accelerate thousands of vessels (each
with the mass of a $\mathrm{km}$ size asteroid) to a much higher time
dilation factor of $\gamma\sim 10^4$. This enables a Type II
civilization to establish a galactic confederation based on a network of
synchronized vessels undergoing such motion on linear trajectories
between nodes (as illustrated in Fig. \ref{Fig:CivNet}). A trip at
$\gamma\sim 10^4$ between nodes separated by $20,000~\mathrm{ly}$ will
take $20$ years in the civilizational frame; a traveler within a galaxy
spanning network of such trajectories can traverse the diameter of
the galaxy in a century as measured in the civilizational frame. 

As the civilization advances further on the Kardashev scale, another
possibility becomes available. A sufficiently powerful civilization can
rearrange stellar-mass black holes within the galaxy so that the spatial
trajectory of an ultrarelativistic vessel is deflected by gravity of the
black holes so as to form a closed path within the galaxy. The advantage
of such an approach is that the vessels do not need to be continuously
accelerated and decelerated, possibly reducing the required maintenance
budget. Moreover, one can extract the rotational energy of a black hole
via the Penrose process \cite{Penrose:1971uk}---a \emph{single} black
hole of typical mass and spin has enough rotational energy to power a
Type II civilization well beyond the age of the universe. Such a
structure provides significant advantages for enhancing the
civilization's longevity and resilience, and can serve as a seat for a
centralized authority for an existing network of linearly accelerating
vessels, turning the aforementioned confederation into a galactic
empire. However, the cost of constructing such a ring is significant,
and requires a time scale on the order of a few million years in the
galactic frame. We find that for a time dilation factor of $10^4$, one
must construct a ring from $6.2\times 10^5$ black holes of mass
$\sim20~M_\odot$. To construct a ring of radius of $10,000~\mathrm{ly}$,
one must move the aforementioned number of black holes a distance of
$200~\mathrm{ly}$, which can be achieved by a sub-Type III civilization
of power capacity $\sim 10^{32}~\mathrm{W}$ in about two million years
in the galactic frame. However, if the civilization is already in a
frame with a time-dilation factor $\gamma\sim 10^4$, this translates to
a construction time of about $\sim300$ years in the civilizational
frame, well-below that of the longest sustained human construction
projects to date.

We have pointed out some implications for the Fermi paradox. Time
dilation effects will of course reduce the frequency of expected visits
to the Earth from a single civilization, but a large number of
extraterrestrial civilizations may compensate for such effects. However,
a highly time-dilated, or redshifted civilization is vulnerable to
destruction by hostile civilizations; placing a large number of objects
in the path of an ultrarelativistic vessel has the potential to cause
catastrophic damage. While this vulnerability can be mitigated to some
degree, potentially hostile civilizations can appear within the
proverbial blink of an eye. In the frame of an approaching vessel with
time dilation factor $\gamma\sim 10^4$, our own human civilization will
appear to have progressed from hunter--gatherer societies (about $12,000$
years ago) to one capable of sending objects into interstellar space
\cite{Voyager_Mission:2025} in about a year. The relative nature of time
dilation in special relativity may in some cases limit a time-dilated
civilization's awareness of an emerging civilization's advancement until
they decelerate to the galactic frame. The possible sudden emergence of
such existential threats to redshifted civilizations provides a strong
motivation for such extraterrestrial civilizations to hide their
presence, and to have some autonomous measures to handle threats as they
emerge. The scenario we consider here in this article provides a strong
motivating factor for the ``Dark Forest''
\cite{Yu:2015JBIS,Hibberd:2025} or ``Berserker'' hypotheses
\cite{Webb:2015}. An interesting question is how such a factor might
enter into a game-theoretic analyses of such scenarios
\cite{Yasser:2020fermi}.

Given the considerations in the preceding paragraph, redshifted
civilizations may be rather difficult to detect. However, the chances of
detection may increase if there is a concentration of civilizations
around supermassive black holes. As the number of civilizations at a
given supermassive black hole increase, orbital lanes become a scarce
resource, and one might expect a higher likelihood for conflict. We have
briefly considered the properties of isotropic, monochromatic signals
from orbits near the photon radius of a black hole, and have found that
the observed signal exhibits a  ``sad trombone'' downward frequency
drift over time, a characteristic feature of fast radio bursts
\cite{CHIME:2019ApJ}. A conflict occurring in orbits near the photon
sphere of supermassive black holes may produce signals with similar
characteristics.

We acknowledge that the considerations in this article may not be
exhaustive; there may be other possibilities beyond our present
imagination satisfying the fundamental constraints listed in the
introduction. We also recognize that future developments may overcome
these constraints. Novel physics and faster-than-light travel are
already addressed in the introduction. Regarding the intolerance of
biological matter to high acceleration, one might consider that
biological matter might eventually be replaced with synthetic
counterparts for the purpose of tolerating higher acceleration
environments. While such a possibility cannot be definitively ruled out,
it is difficult to establish meaningful constraints without detailed
speculation about future technologies (though we note that
postbiological evolution of civilization has been considered in the
literature \cite{Dick:2003,Cirkovic:2005rt,Dick:2008postbiological}).
Moreover, it is unclear whether a synthetic counterpart that can be
engineered to tolerate any environment or circumstance without pain or
discomfort can still be regarded as human---it is rather difficult for
human authors (as is the case for us) to evaluate such a possibility at
present. As we do not wish  in this article to enter into unconstrained
speculation (or be forced into a discussion of philosophical or moral
implications), we leave the consideration of such a scenario for future
work.

As we reiterated, we have refrained from speculating on the full scope
of technological developments needed to create the infrastructure
required for a civilization to migrate to a highly time-dilated frame;
technological developments in the distant future may be beyond our
present ability to foresee. It is certainly possible that there exists
some insurmountable technological obstacle preventing the realization of
redshifted civilizations, but this is well beyond the scope of our work.
Our aim here is merely to demonstrate that general relativity and
biological limits present no fundamental physics obstacle to the
migration of human civilization to a redshifted frame. Nevertheless, in
the remainder of this article, we make an attempt to speculate on the
necessary technological capacities that human civilization must develop
to migrate to a time-dilated frame.

As pointed out in \cite{Sagan:1963} it is unlikely that extreme Lorentz
factors on the order of $10^4$ can be reached by way of rocket
propulsion. A calculation using the relativistic generalization
\cite{Forward:1995,Sanger:1953} of the Tsiolkovsky rocket equation
\cite{Blagonravov:1965,Tsiolkovsky:1903,Maksimov:2007,Moore:1811}
indicates that a rather extreme mass ratio will be required to reach the
Lorentz factors required. Proposals like the Bussard ramjets
\cite{Bussard:1960,Fishback:1968} are limited by drag (and a recent
study \cite{Schattschneider:2022} suggests that the construction of a
Bussard ramjet is infeasible even for a Type II civilization). However,
these limitations do not apply to direct-impulse propulsion, such as
beamed propulsion
\cite{Marx:1966,Landis:1997,Landis:2001,Lubin:2015,Forward:1984}. Such
technology is already being considered for the Breakthrough Starshot
initiative \cite{Guillochon:2015seti,Lubin:2016roadmap,Lubin:2024large},
which aims to send a fleet of probes to Alpha Centauri within the span
of a human lifetime \cite{Parkin:2018breakthrough}. We also note that
the Large Hadron Collider (LHC) can already accelerate protons to a
Lorentz factor greater than $10^3$ (based on a beam energy of
$6.5~\mathrm{TeV}$ for the 2015 proton run \cite{Bruce:2016iew}), and
the proposed Future Circular Collider can achieve Lorentz factors
exceeding $10^4$ \cite{Benedikt:2020ejr}. Whether macroscopic objects
can be accelerated to such Lorentz factors remains an open question
which we leave for future consideration. 

Perhaps the most basic and significant technical hurdle to overcome,
should humanity choose to embark on the migration to a time-dilated
frame, is the significant increase in civilizational power capacity to
the scales required for such an undertaking. In our view, a
civilization's capabilities are determined mainly by raw power capacity,
regardless of the specific technologies they possess. Currently, the
power capacity of human civilization is increasing by 1\% annually,
indicating that human civilization will reach the Type I threshold in
less than a millennium, and possibly Type II in roughly three millennia.
However, if one presumes that global power capacity follows the growth
of the population \cite{Ehrlich:1971impact,Syvitski:2020extraordinary},
one might worry that the power capacity may plateau when the population
peaks sometime in the 21st century (according to current projections
\cite{UNDESA:2024world}). In order to sustain and perhaps accelerate the
power capacity of human civilization in spite of these projections, it
may be necessary to develop autonomous self-replicating von Neumann
machines \cite{vonNeumann:1966theory,Sipper:1998fifty}, as suggested in
\cite{Tipler:1980}. Such machines need not be mechanical in nature; as
of this writing, biological photosynthesis exceeds the power capacity of
human civilization by an order of magnitude \cite{Nealson:1999life}, and
it is known that photosynthesizing cyanobacteria can have a doubling
time on the order of hours
\cite{Kondo:1997circadian,Mori:1996circadian}. In our view, the
development of controllable self-replicating machines (whether they be
biological or otherwise \cite{Davies:2010book,Sipper:1998fifty}), and
methods for managing the associated risks
\cite{Wang:2019synthetic,Li:2021advances,Freitas:2000some,Freitas:2006molecular}
will be critical steps in human civilization's further advancement on
the Kardashev scale.

%=======================================================================

%-----------------------------------------------------------------------
%-----------------------------------
%-----------------
%--------
%---
%-
%
%
%-
%---
%--------
%-----------------
%-----------------------------------
%-----------------------------------------------------------------------

%=======================================================================
%		ACKNOWLEDGMENTS
%=======================================================================

\section{Acknowledgments}
This article is dedicated to the memory of Pavel Bakala, who mentored
one of us (C. R.) during the early stages of this project. C. R. thanks
Avi Loeb for comments on an earlier version of this work. J. C. F. is
supported by the European Union and Czech Ministry of Education, Youth
and Sports through the FORTE project No.
CZ.02.01.01/00/22{\_}008/0004632.

%=======================================================================

%-----------------------------------------------------------------------
%-----------------------------------
%-----------------
%--------
%---
%-
%
%
%-
%---
%--------
%-----------------
%-----------------------------------
%-----------------------------------------------------------------------

%=======================================================================
%-----------------------------------------------------------------------
%
%		APPENDIX
%
%-----------------------------------------------------------------------
%=======================================================================
\appendix

\section{Special and general relativistic
considerations}\label{sec:AppendixGRSR}

In this section, we review some standard results in special and general
relativity, which can be found in the textbooks
\cite{TaylorWheeler:1992,Carroll,Poisson,Wald,Weinberg,MTW}. Here, we
follow the sign conventions of Misner, Thorne, and Wheeler \cite{MTW}.

\subsection{Motion in special relativity}
We begin by reviewing motion in special relativity. Here, space and time
are described as a spacetime manifold with coordinates $x^\mu$, where
Greek superscripts and subscripts in the set of symbols
$\{\mu,\nu,\varrho,\sigma\}$ take values from the set $\{0,1,2,3\}$,
with the value $0$ indicating the time coordinate $t$---that is, we set
$x^0=c t$ (with $c=299,792,458~\mathrm{m}/\mathrm{s}$ being the speed of
light). Distances and proper times are measured according to the
Minkowski metric $\eta_{\mu \nu }=\mathrm{diag}(-1,1,1,1)$, which, for
spatial Cartesian coordinates $(\ubar{x},\ubar{y},\ubar{z})$ may be
written:
\begin{equation} \label{EqA:LineElement}
    ds^2=\eta_{\mu \nu} dx^\mu dx^\nu
    =-c^2dt^2+d\ubar{x}^2+d\ubar{y}^2+d\ubar{z}^2,
\end{equation}
where $s$ is an arc length parameter and Einstein summation convention
is applied to the indices $\mu$ and $\nu$---we shall continue to employ
Einstein summation convention (in which pairs of indices with the same
symbol are summed over) on the spacetime indices
$\{\mu,\nu,\varrho,\sigma\}$ unless otherwise stated. The motion of a
particle with mass can be parameterized by proper time $\tau$ (the time
as measured by a clock attached to the particle), so that the coordinate
position of the particle may be written as $x^\mu=X^\mu(\tau)$. The
derivative of the particle's spacetime coordinates with respect to
$\tau$ is its four-velocity $U^\mu$:
\begin{equation} \label{EqA:Four-Velocity}
    U^\mu:=\frac{dX^\mu}{d\tau}.
\end{equation}
Now the metric $\eta_{\mu\nu}$ defines an inner product; for vectors
$U^\mu$ and $V^\mu$, one may write $\langle U,V\rangle=\eta_{\mu
\nu}U^\mu V^\nu$. The norm squared is then $\langle U,U\rangle=\eta_{\mu
\nu}U^\mu U^\nu=-c^2$ under the condition that displacements in proper
time correspond to arc length: $c|d\tau|=|ds|$. It is convenient to
define the time coordinate of the four-velocity $U^\mu$ as:
\begin{equation} \label{EqA:Four-Velocity-TimeComponent}
    \gamma:=\frac{U^0}{c}=\frac{dt}{d\tau}=\frac{1}{\sqrt{1-(v/c)^2}},
\end{equation}
where the last equality can be obtained from $\eta_{\mu \nu} U^\mu
U^\nu=-c^2$ and an application of the chain rule, with the speed of the
particle $v$ given by:
\begin{equation} \label{EqA:Three-Vel}
    \begin{aligned}
    v:=\sqrt{\left(\frac{d\ubar{x}}{dt}\right)^2
             +\left(\frac{d\ubar{y}}{dt}\right)^2
             +\left(\frac{d\ubar{z}}{dt}\right)^2}.
    \end{aligned}
\end{equation}
One might recognize $\gamma$ as the Lorentz factor, which provides a
measure of time dilation for special relativistic motion. From the
four-velocity, one can construct the four-momentum $p^\mu$:
\begin{equation} \label{EqA:Four-Momentum}
    p^\mu:=m U^\mu,
\end{equation}
which satisfies the expression $\eta_{\mu \nu }p^\mu p^\nu=-m^2 c^2$.
The time component can be interpreted as an energy
\begin{equation} \label{EqA:Energy}
    E:=p^0 c=m c \sqrt{c^2+\gamma^2 v^2},
\end{equation}
which yields the famous formula $E=mc^2$ in the limit $v\rightarrow0$.

The four-acceleration $A^\mu$ is given by:
\begin{equation} \label{EqA:Four-Acceleration}
    A^\mu:=\frac{d^2X^\mu}{d\tau^2}=\frac{dU^\mu}{d\tau}
    =\frac{1}{m}\frac{dp^\mu}{d\tau}.
\end{equation}
It is not too difficult to show from $\eta_{\mu \nu }U^\mu U^\nu=-c^2$
that the four-acceleration satisfies:
\begin{equation} \label{EqA:Four-Acceleration-Orth}
    \langle A,U\rangle= \eta_{\mu\nu} A^\mu U^\nu = 0,
\end{equation}
so that the four-acceleration $A^\mu$ is orthogonal to the four-velocity
$U^\mu$. The component $A^0$ may be interpreted in terms of the power
required to accelerate the particle:
\begin{equation} \label{EqA:Power}
    P:=\frac{d E}{dt}
      =\frac{c}{\gamma}\frac{dp^0}{d\tau}=\frac{m c A^0}{\gamma}.
\end{equation}

We briefly describe here two types of accelerated motion. The first is
circular motion, which may be described by the following:
\begin{equation} \label{EqA:CircularSR}
    X_\mathrm{\circ}^\mu(\tau) 
        = \left( 
                 c \gamma_\mathrm{\circ} \tau,
                 R_\mathrm{\circ} \cos(\omega_\mathrm{\circ}\tau),
                 R_\mathrm{\circ} \sin(\omega_\mathrm{\circ}\tau),
                 0 \right),
\end{equation}
where $R_\mathrm{\circ}$ is the radius of the circular motion,
$\gamma_\mathrm{\circ}$ is the Lorentz factor, and
$\omega_\mathrm{\circ}$ is the orbital frequency as measured by an
observer undergoing the motion. Applying Eqs. \eqref{EqA:Four-Velocity}
and \eqref{EqA:Four-Acceleration} to obtain the four-velocity
$U_\mathrm{\circ}^\mu$ and four-acceleration $A_\mathrm{\circ}^\mu$, one
can obtain two expressions from $\eta_{\mu \nu }U_\mathrm{\circ}^\mu
U_\mathrm{\circ}^\nu=-c^2$ and
$a_\mathrm{\circ}=\sqrt{\eta_{\mu\nu}A_\mathrm{\circ}^\mu
A_\mathrm{\circ}^\nu}$ (with $a_\mathrm{\circ}$ being the magnitude of
the acceleration experienced by particles following Eq.
\eqref{EqA:CircularSR}):
\begin{equation} \label{EqA:CircularSRrels}
    R_\mathrm{\circ} = \frac{c^2 (\gamma_\mathrm{\circ}^2-1)}
                            {a_\mathrm{\circ}},
    \qquad
    \omega_\mathrm{\circ} = \frac{a_\mathrm{\circ}}
                                 {c \sqrt{\gamma_\mathrm{\circ}^2-1}}.
\end{equation}
Next, we consider the motion of a uniformly accelerating particle moving
in the $\ubar{x}$ direction (and starting at the origin $X^\mu(0)$). The
motion may be parametrically written as \cite{Sagan:1963} (cf. Eq.
(9.126) in \cite{Carroll}):
\begin{equation} \label{EqA:Hyperbolic}
    X^\mu(\tau) = \left(\frac{c^2 \sinh(a \tau/c)}{a},
                    \frac{c^2 (\cosh(a \tau/c)-1)}{a},0,0\right).
\end{equation}
where $a:=\sqrt{\eta_{\mu\nu}A^\mu A^\nu}$ is the magnitude for the
acceleration experienced by the particle in its own instantaneous rest
frame, defined as the reference frame momentarily aligned with the
four-velocity $U^\mu$.

\subsection{Motion in the Schwarzschild spacetime}
\subsubsection{Invariants}
We now consider the motion of massive free falling particles in the
Schwarzschild spacetime, which describes the gravitational field of a
spherically symmetric object in a vacuum. In the curved spacetime
geometries of general relativity, the motion of a massive free falling
particle is a geodesic, or a curve that maximizes the proper time of the
spacetime curve $x^\mu=X^\mu(\tau)$. The properties of geodesics in the
Schwarzschild spacetime are well-known, and we summarize them here. 

In spherical coordinates $(t,r,\theta,\phi)$, the Schwarzschild
spacetime is described by the line element:
\begin{equation}\label{EqA:Schwarzschild}
    ds^2 = g_{\mu \nu} dx^\mu dx^\nu
         = - f(r) c^2 dt^2 + dr^2/f(r) + r^2 d\Omega^2,
\end{equation}
where $f(r):=1-2M/r$ and $d\Omega^2:=d\theta^2+(\sin(\theta) \,
d\phi)^2$, with the mass parameter $M:=G \bar{M}/c^2$ has the value of
half the Schwarzschild radius $r_\mathrm{s}:=2G \bar{M}/c^2$, where
$\bar{M}$ is the mass of the spherically symmetric object. As before,
the metric components $g_{\mu\nu}$ define an inner product, and can be
read off from the line element Eq. \eqref{EqA:Schwarzschild} to obtain
$g_{00}=-f(r)$, $g_{11}=1/f(r)$, $g_{22}=r^2$, and $g_{33}=r^2 \sin^2
\theta$, with all other components vanishing.

As before, we consider a curve $x^\mu=X^\mu(\tau)$ parameterized by
proper time $\tau$, with the respective four-velocity $U^\mu$ defined as
in Eq. \eqref{EqA:Four-Velocity}. Since the metric $g_{\mu \nu}$ is
independent of the time coordinate $t$ and the angular coordinate
$\theta$, one may obtain the geodesic invariants (cf. Sec. 1.5 of
\cite{Poisson}):
\begin{equation}\label{EqA:InvariantQuantities}
    \begin{aligned}
    e &= -g_{0 \mu} U^\mu = c \dot{t} f(r), \\
    l &= g_{3 \mu} U^\mu = r^2\dot{\phi},
    \end{aligned}
\end{equation}
where $\dot{t}:=\tfrac{dt(\tau)}{d\tau}$ and
$\dot{\phi}:=\tfrac{d\phi(\tau)}{d\tau}$. The quantities $e$ and $l$ are
the respective specific (per unit mass) energy and angular momentum of a
free-falling particle in the Schwarzschild spacetime, and remain
constant along the geodesic. The four-velocity may then be written as:
\begin{equation}\label{EqA:FourVelocity}
    U^\mu = \left(-e/f(r),\dot{r},0,l/r^2\right).
\end{equation}

Without loss of generality, we restrict the motion to the equatorial
plane $\theta=\pi/2$. For massive particles, the norm of the four
velocity satisfies $\eta_{\mu \nu }U^\mu U^\nu=-c^2$, which may be
rewritten as:
\begin{equation}\label{EqA:Norm4v}
    \begin{aligned}
    \left(\frac{dr}{d\tau}\right)^2 
    = e^2-f(r)\left(c^2+\frac{l^2}{r^2}\right),
    \end{aligned}
\end{equation}
For our purposes, it is sufficient to consider a radius $r=r_\mathrm{c}$
defined as the radius along the geodesic satisfying
$\dot{r}:={dr}/{d\tau}=0$, which corresponds to either circular orbits
or to the point of closest approach along the trajectory. Under this
condition, the energy may be written:
\begin{equation}\label{EqA:SpecEnergy}
    \begin{aligned}
    e^2 = f(r_\mathrm{c})\left(c^2+{l^2}/{r_\mathrm{c}^2}\right).
    \end{aligned}
\end{equation}

It is perhaps worth noting that the energy invariant is closely related
to the time dilation factor; from Eq. \eqref{EqA:InvariantQuantities},
one can obtain the following expression:
\begin{equation}\label{EqA:GammaFactorGeneral}
    \Gamma:=\frac{dt}{d\tau}
           =\dot{t}=\frac{e}{c f(r)}.
\end{equation}
Observe also that at large radii, $f(r)\rightarrow 1$, so that in this
limit $\Gamma\approx e/c$.

\subsubsection{Frames}

For later convenience, we construct a local reference frame adapted to
$U^\mu$ at a radius $r_\mathrm{c}$ in which $\dot{r}=0$. From $U^\mu$,
one can construct a timelike vector of the form:
\begin{equation}\label{EqA:TetradTime}
    \begin{aligned}
    \hat{e}^{\mu}_{t}
        & = c^{-1}\left(-e/f(r_\mathrm{c}),0,0,
                    l/r_\mathrm{c}^2\right).
    \end{aligned}
\end{equation}
The spatial directions are determined by the frame vectors:
\begin{equation}\label{EqA:Tetrad}
    \begin{aligned}
    \hat{e}^{\mu}_{\phi}
        & = c^{-1}\left(l/\sqrt{r_\mathrm{c}^2 f(r_\mathrm{c})},0,0,
                    -\sqrt{l^2+c^2r_\mathrm{c}^2}/r_\mathrm{c}^2\right),
                    \\
    \hat{e}^{\mu}_{\theta} & = \left(0,0,1/r_\mathrm{c},0\right),
    \qquad 
    \hat{e}^{\mu}_{r} = \left(0,\sqrt{f(r_\mathrm{c})},0,0\right).
    \end{aligned}
\end{equation}
One can verify that these vectors are orthonormal to each other, and
orthogonal to $\hat{e}^{\mu}_{t}$, in particular that they satisfy the
expressions $g_{\mu\nu}\hat{e}^{\mu}_{i}\hat{e}^{\nu}_{j}=\delta_{ij}$
and $g_{\mu\nu}\hat{e}^{\mu}_{i}U^{\nu}=0$, where $\delta_{ij}$ is the
Kronecker delta and the indices $i,j\in\{1,2,3\}$ correspond to the
spatial coordinates $(r,\theta,\phi)$.

\subsubsection{Circular orbits}

We now consider circular orbits. From Eq.
\eqref{EqA:InvariantQuantities} we identify an effective potential (cf.
Eq. (6.3.15) in \cite{Wald}):
\begin{equation}\label{EqA:EffPot}
    V(r) = \frac{f(r)}{2}\left(c^2+\frac{l^2}{r^2}\right).
\end{equation}
Circular orbits occur when the orbital radius $r_\mathrm{c}$ satisfies
the conditions $V(r_\mathrm{c})=e$ and $V'(r_\mathrm{c})=0$. The first
condition establishes a relationship between $r_\mathrm{c}$ and the
specific energy $e$, and the second condition relates $r_\mathrm{c}$ to
the specific angular momentum $l$:
\begin{equation}\label{EqA:AngMomCirc}
    \begin{aligned}
    l^2 = \frac{M r_\mathrm{c}^2}{r_\mathrm{c}-3M} 
    \quad\Leftrightarrow\quad
    r_\mathrm{c} = \frac{l(l\pm\sqrt{l^2-12 M^2})}{2M}.
    \end{aligned}
\end{equation}
Equation \eqref{EqA:AngMomCirc} and the expression for $e$ in Eq.
\eqref{EqA:SpecEnergy} may be combined with Eq.
\eqref{EqA:GammaFactorGeneral} to obtain an expression for $\dot{t}$:
\begin{equation}\label{EqA:GammaFactorCirc}
    \Gamma:=\frac{dt}{d\tau}
           =\sqrt{\frac{r_\mathrm{c}}{r_\mathrm{c}-3M}}.
\end{equation}
This establishes the relative time dilation factor between observers
following the orbital trajectory at $r_\mathrm{c}$ and static observers
at $r=\infty$. Note that in the limit $r_\mathrm{c}\rightarrow 3M$
(corresponding to the photon radius), the time dilation factor $\Gamma$
diverges.  

It is well-known that at the photon radius $r_\mathrm{c}\sim 3M$ are
unstable, so it is perhaps worth discussing the timescale for the
destabilization of circular orbits close to $r_\mathrm{c}\sim 3M$. This
timescale is characterized by the Lyapunov exponent, which is given by
the second derivative of the effective potential \cite{Cardoso:2008bp}:
\begin{equation}\label{EqA:LyapunovExponent}
    \lambda=c\frac{\sqrt{-V''(r_\mathrm{c})}}{\Gamma}
           =\frac{c\sqrt{2M(6M-r_\mathrm{c})}}{r_\mathrm{c}^2}.
\end{equation}
The destabilization timescale $t_0$ is the inverse of the Lyapunov
exponent, $t_0=1/\lambda$. Near the photon radius $r_\mathrm{c}\sim 3M$,
one finds $t_0\approx3\sqrt{3}G\bar{M}/c^3$. A calculation of the
Lyapunov exponent with the parameters of Fig. \ref{Fig:GammavMass} with
the mass of Sgr A* yields an instability timescale of
$1/\lambda\sim78~\mathrm{s}$.

\subsubsection{Scattering trajectories}

Scattering problems are characterized by an impact parameter. Following
\cite{Wald}, we introduce an \emph{apparent} impact parameter, which may
be defined as $b:=l/e$ (this coincides with the usual impact parameter
in the limit $e\gg c^2$). Combined with \eqref{EqA:SpecEnergy}, this
definition yields:
\begin{equation}\label{EqA:ImpactPar}
    \begin{aligned}
    b:=\frac{l}{e}
      = r_\mathrm{c}
        \sqrt{\frac{r_\mathrm{c}}{r_\mathrm{c}-2M}-\frac{c^2}{e^2}},
    \end{aligned}
\end{equation}
where here, $r_\mathrm{c}$ is the point of closest approach. One may
observe that in the limit $e\gg c$ and $r_\mathrm{c}\gg M$, $b\sim
r_\mathrm{c}$. Given Eq. \eqref{EqA:ImpactPar} and the expression for
$l$ in Eq. \eqref{EqA:InvariantQuantities}:
\begin{equation}\label{EqA:ODEScattering}
    \frac{dr}{d\phi} = \frac{\dot r}{\dot\phi} 
    = \frac{r\sqrt{e^2 r^2-f(r)(l^2+r^2 c^2)}}{l},
\end{equation}
which one may integrate to obtain an expression for the total change in
angular coordinate over the entire course of the scattering trajectory:
\begin{equation}\label{EqA:Integral}
    \begin{aligned}
    \Delta\phi
    &=2\int_{r_\mathrm{c}}^\infty
       \frac{b e ~ dr}{\sqrt{e^2 r(r^3-b^2(r-2M))-c^2 r^3 (r-2M)}},\\
    &=\pi+\frac{2M(2 e^2-c^2)}{r_\mathrm{c}(e^2-c^2)}+\mathscr{O}(M^2),
    \end{aligned}
\end{equation}
where the second line is obtained by expanding the integrand to first
order in $M$. The deflection angle $\Delta\psi:=\Delta\phi-\pi$ is given
by:
\begin{equation}\label{EqA:DeflectionAngle}
    \Delta\psi
    =
    \frac{2M(2 e^2-c^2)}{r_\mathrm{c}(e^2-c^2)}+\mathscr{O}(M^2).
\end{equation}
In the high energy limit $e\gg c$, this expression reduces to
$\Delta\psi\approx 4M/r_\mathrm{c}$. We remind the reader that for $e\gg
c$ and $r_\mathrm{c}\gg M$, $b\sim r_\mathrm{c}$, so that we recover the
well-known light deflection formula $\Delta\psi\approx 4M/b$.

\subsubsection{Photon trajectories}

Here, we consider photon trajectories, which are described by curves
$X^\mu(\lambda)$ parameterized by an affine parameter $\lambda$, with a
tangent vector $k^\mu:=dX^\mu/d\lambda$ satisfying the null condition:
\begin{equation}\label{EqA:NullCondition}
    g_{\mu \nu} k^\mu k^\nu = 0.
\end{equation}
We are interested in signals emitted from circular orbits of radius
$r_\mathrm{c}$, and for this reason it will be useful to write $k^\mu$
in the basis of the frame vectors:
\begin{equation}\label{EqA:NullvecOrthBasis}
\begin{aligned}
    \left. k^\mu \right|_{r=r_\mathrm{c}} = \nu_\mathrm{s} 
    (& \hat{e}^{\mu}_{t}
        +
        \sin(\vartheta_1) \cos(\vartheta_2)\hat{e}^{\mu}_{r} 
        + 
        \sin(\vartheta_1) \sin(\vartheta_2) \hat{e}^{\mu}_{\theta}\\
     &+ \cos(\vartheta_1)\hat{e}^{\mu}_{\phi}).
\end{aligned}
\end{equation}
where $\nu_\mathrm{s}$ is the source frequency factor, $\vartheta_1$ is
the angle that the spatial part of $k^\mu$ makes with
$\hat{e}^{\mu}_{\phi}$, and $\vartheta_2$ the angle made with
$\hat{e}^{\mu}_{r}$.

Photons follow null geodesics, and one can define invariants for photon
trajectories in the same manner as in \eqref{EqA:InvariantQuantities}.
It is convenient to regard the energy invariant as the observed
frequency factor $\nu_\mathrm{o}$, which corresponds (up to some
constant factor) to the frequency of the photon when it reaches an
observer at infinity. One may write:
\begin{equation}\label{EqA:PhotonFrequencyFactor}
    \nu_\mathrm{o}(\vartheta_1) = -g_{0 \mu} k^\mu 
    = \frac{\nu_\mathrm{s} \sqrt{f(r_\mathrm{c})} 
    \left[  \sqrt{r_\mathrm{c}f(r_\mathrm{c})} 
            \mp 
            \sqrt{M} \cos (\vartheta_1)\right]}
            {\sqrt{(r_\mathrm{c}-3 M)}},
\end{equation}
where the sign $\pm$ corresponds to the sign of $l$ and the second
equality is obtained from Eq. \eqref{EqA:NullvecOrthBasis} and Eq.
\eqref{EqA:AngMomCirc}. Another invariant is the apparent impact
parameter:
\begin{equation}\label{EqA:PhotonImpactParam}
    B(\vartheta) 
    = \frac{g_{3 \mu} k^\mu}{\nu_\mathrm{o} } 
    = \frac{\pm \sqrt{M}-\cos (\vartheta)\sqrt{r_\mathrm{c}-2M} }
        {\sqrt{r_\mathrm{c}-2M}\mp\sqrt{M}\cos (\vartheta)}
        \frac{r_\mathrm{c}}{\sqrt{f(r_\mathrm{c})}}.
\end{equation}
In the case of null geodesics confined to the equatorial plane
$\theta=\pi/2$ (setting $\vartheta_2=0$), one can relate the angle
$\vartheta_1$ to the observer position angle $\phi_\mathrm{o}$ via the
integral (cf. Eq. 6.3.38 in \cite{Wald}):
\begin{equation}\label{EqA:IntegralNull}
    \begin{aligned}
    \phi_\mathrm{o}(\vartheta_1)
    &=\int_{r_\mathrm{c}}^\infty
       \frac{B(\vartheta_1) ~ dr}{\sqrt{r(r^3-B(\vartheta_1)^2(r-2M))}}.
    \end{aligned}
\end{equation}
If one can evaluate the above integral, then one can parametrically
obtain the dependence of the observed frequency $\nu_\mathrm{o}$ on the
observer angle $\phi_\mathrm{o}$, assuming the emitted signal is
isotropic and monochromatic. In particular, one can plot
$\nu_\mathrm{o}(\phi_\mathrm{o})$ by treating $\vartheta_1$ as a
parameter.

\subsection{Tidal accelerations}
Tidal accelerations experienced by observers on circular equatorial
orbits in the more general Kerr spacetime are explicitly described in
\cite{Marck:1983,Fishbone:1973ApJ}, but for the reader's convenience, we
rederive these results in the simpler setting of the Schwarzschild
spacetime. 

In the vicinity of a geodesic, the tidal accelerations due to the
curvature of spacetime are given by the geodesic deviation equation
\cite{MTW,Pirani:1956tn,Synge:1934zza} (cf. Sec. 1.10 of
\cite{Poisson}):
\begin{equation}\label{EqA:GeodesicDeviation}
    A_\chi^\mu:= \frac{D^2 \chi^\mu}{d\tau^2}
    =
    R{^\mu}_{\xi\sigma\nu}U^\xi U^\sigma \chi^\nu,
\end{equation}
where $\chi^\nu=\chi^\nu(\tau)$ is the separation vector between a
geodesic parameterized as $x^\mu(\tau)$ and a neighboring geodesic
parameterized as $x'^\mu(\tau)=x^\mu(\tau)+\chi^\mu(\tau)$. The formula
is interpreted as describing the relative acceleration $A_\chi^\mu$
between the neighboring geodesics $x^\mu(\tau)$ and $x'^\mu(\tau)$. It
is convenient to choose a separation vector $x^\mu(\tau)$ that is
spacelike and orthogonal to $U^\mu$. The separation vector can then be
decomposed in a basis the frame vectors defined in Eq.
\eqref{EqA:Tetrad}:
\begin{equation}\label{EqA:SepVecDecomp}
    \chi^\mu=
         \hat{\chi}^r \hat{e}^{\mu}_{r}
        +\hat{\chi}^\theta \hat{e}^{\mu}_{\theta}
        +\hat{\chi}^\phi \hat{e}^{\mu}_{\phi},
\end{equation}
where $\hat{\chi}^r$, $\hat{\chi}^\theta$, and $\hat{\chi}^\phi$ are the
components of $\chi^\mu$ (which have units of length) in the orthonormal
basis of vectors $\hat{e}^{\mu}_{i}$. Inserting the above expression
into Eq. \eqref{EqA:GeodesicDeviation} yields the following expressions
(assuming only $\dot{r}=0$) for the components of the acceleration in
the orthonormal basis:
\begin{equation}\label{EqA:AccelComponentsGen}
    \begin{aligned}
    \hat{A}^r &:= g_{\mu\nu}A^\mu\hat{e}^{\nu}_{r}
    =\frac{M (3l^2+2c^2 r^2)}{r^5}\hat{\chi}^r,\\
    \hat{A}^\theta &:= g_{\mu\nu}A^\mu\hat{e}^{\nu}_{\theta}
    =-\frac{M (3l^2+c^2 r^2)}{r^5}\hat{\chi}^\theta,\\
    \hat{A}^\phi &:= g_{\mu\nu}A^\mu\hat{e}^{\nu}_{\phi}
    =-\frac{c^2 M}{r^3}\hat{\chi}^\phi.
    \end{aligned}
\end{equation}
Up to a difference in sign convention and a frame rotation, the above
expressions agree with those obtained from the tidal tensor in
\cite{Marck:1983}. The longitudinal component $\hat{A}^\phi$ is finite
and bounded so long as one remains at a finite distance from the horizon
radius $2M$---in particular $\hat{A}^\phi$ has a finite magnitude at the
photon radius $r=3M$, and is in fact suppressed for large $M$. The
components $\hat{A}^r$ and $\hat{A}^\theta$ transverse to the orbital
direction can become large; in the case of circular orbits, one can see
from Eq. \eqref{EqA:AngMomCirc} that the angular momentum $l$ can become
large for circular orbit radii $r_\mathrm{c}$ that approach the photon
radius $3M$.

For circular orbits, one may expand the transverse components
$\hat{A}^r$ and $\hat{A}^\theta$ in the parameter
$\rho_{\mathrm{circ}}:=r_\mathrm{c}-3M$, that is, the coordinate
difference between the orbital radius $r_\mathrm{c}$ and the photon
radius $3M$. We obtain:
\begin{equation}\label{EqA:AccelComponentsCirc}
    \begin{aligned}
    \hat{A}^r_{\mathrm{circ}} &= 
        \alpha_{\mathrm{circ}}\hat{\chi}^r 
        + \mathscr{O}(\rho_{\mathrm{circ}}^0),\\
    \hat{A}^\theta_{\mathrm{circ}} &= 
        -\alpha_{\mathrm{circ}}\hat{\chi}^\theta 
        + \mathscr{O}(\rho_{\mathrm{circ}}^0),
    \end{aligned}
\end{equation}
where we have defined:
\begin{equation}\label{EqA:AccelComponentsCircDiv}
    \alpha_{\mathrm{circ}} 
    :=
    \frac{c^2}{9M\rho_{\mathrm{circ}}}
    =
    \frac{c^4}{9 G \bar{M}\rho_{\mathrm{circ}}}.
\end{equation}

At the point of closest approach in a scattering trajectory, one may set
$l=be$ and (after making use of Eq. \eqref{EqA:ImpactPar}) expand in
$1/e$:
\begin{equation}\label{EqA:AccelComponentsSc}
    \begin{aligned}
    \hat{A}^r_{\mathrm{sc}} &= 
        \alpha_{\mathrm{sc}}\hat{\chi}^r + \mathscr{O}(1/e^0),\\
    \hat{A}^\theta_{\mathrm{sc}} &= 
        -\alpha_{\mathrm{sc}}\hat{\chi}^\theta + \mathscr{O}(1/e^0),
    \end{aligned}
\end{equation}
where we have defined:
\begin{equation}\label{EqA:AccelComponentsScDiv}
    \alpha_{\mathrm{sc}} :=
        \frac{3 M e^2}{(r_\mathrm{c}-2M)r_\mathrm{c}^2}
    = \frac{3 G \bar{M} c^2 \Gamma_\infty^2}
           {(c^2 r_\mathrm{c}-2 G\bar{M})r_\mathrm{c}^2},
\end{equation}
where Eq. \eqref{EqA:GammaFactorGeneral} is employed in the second
(approximate) equality, with $\Gamma_\infty$ (evaluated at
$f(r=\infty)=1$) coinciding with the special relativistic factor
$\gamma$ in the limit $r\rightarrow \infty$. Equations
\eqref{EqA:AccelComponentsCircDiv} and \eqref{EqA:AccelComponentsScDiv}
provide estimates for the maximum transverse acceleration components
across a unit distance, as experienced by observers traveling along a
geodesic in the Schwarzschild spacetime.

%=======================================================================

%-----------------------------------------------------------------------
%-----------------------------------
%-----------------
%--------
%---
%-
%
%
%-
%---
%--------
%-----------------
%-----------------------------------
%-----------------------------------------------------------------------

%=======================================================================
%		BIBLIOGRAPHY
%=======================================================================

\end{document}